\documentclass[%
 reprint,
 amsmath,amssymb,
 aps,
 prx,
floatfix,
]{revtex4-2}

\usepackage{graphicx}
\usepackage{dcolumn}
\usepackage{bm}
\usepackage{chemformula}
\usepackage{bbold}
\usepackage{dsfont}
\usepackage{xcolor}
\definecolor{costumgreen}{rgb}{0, 0.91, 0.047} 
\definecolor{costumred}{rgb}{1, 0, 0.0625}
\definecolor{costumblue}{rgb}{0, 0.238, 0.918}

\usepackage{xr-hyper}
\usepackage[linesnumbered,ruled,vlined]{algorithm2e}
\usepackage{algpseudocode}
\usepackage{amssymb}
\usepackage{xspace}
\usepackage{caption}
\usepackage{subcaption}
\usepackage{hyperref}

\usepackage{amsthm}

\newcommand{\new}[1]{{#1}}

\begin{document}

\preprint{???}

\title{Emergence of generic first-passage time distributions\\
for large Markovian networks}

\author{Julian B. Voits\textsuperscript{1,2}}
\author{Ulrich S. Schwarz\textsuperscript{1,2}}%
 \email{Corresponding author: schwarz@thphys.uni-heidelberg.de}
\affiliation{%
\textsuperscript{1}Institute for Theoretical Physics, University of Heidelberg, Germany\\ \textsuperscript{2}BioQuant-Center for Quantitative Biology, University of Heidelberg, Germany
}%




\date{\today}

\begin{abstract}
First-passage times are often the most relevant aspect of a complex Markovian network because they signify when information processing has resulted in a definite decision. Previous studies have shown that for kinetic proofreading networks in the limit of large network size the first-passage time distribution converges either to a delta or to an exponential distribution. Remarkably, these two forms correspond to the two extreme distributions of minimal and maximal entropy for a fixed mean, respectively. Here we build on the connection between first-passage times and graph theory to show that these two limits are not model-specific, but arise generically in Markovian networks from the distribution of the eigenvalues of the generator matrix. A deterministic peak emerges when infinitely many eigenvalues contribute, while the exponential limit arises from a single dominant eigenvalue. We also show that the exponential limit emerges robustly for reversible networks when \new{the mean first-passage time from the initial state to the target state becomes much larger than the mean first-passage time in the reverse direction}. In contrast, the deterministic limit is not obtained from a simple reversal of this condition, but \new{follows from a non-vanishing conductance or a mean-residual lifetime of the process which becomes small compared to the mean first-passage time in the long-time limit}. This reveals a fundamental asymmetry between the two regimes. Our theoretical analysis is illustrated and validated by computer simulations of one-step master equations and random networks.
\end{abstract}

\maketitle


\section{Introduction}

Although real world networks often show an overwhelming
level of complexity \cite{boccaletti2006complex,newman2011structure,newman2018networks}, 
often it matters mainly when a
certain absorbing state is reached, 
namely when this state marks
the completion of a certain process \cite{hinrichsen2000non,baronchelli2006sharp,redner2001guide,condamin2007first,metzler2014first}. 
This is especially true in biological systems,
which often have to switch to another program
once a certain state is reached. While in development
this would be the completion of the organism, 
in cellular signaling it could be the decision
to respond to an extracellular signal \cite{tkavcik2025information}. A prominent
example of the latter case would be kinetic proofreading,
a process in which the accuracy of the decision-making
process is increased by energy consumption \cite{hopfield1974kinetic}. Possible
outcomes include cell division in response to
growth factors or activation of the immune system
after detecting a pathogen \cite{ninio1975kinetic,mckeithan1995kinetic}.

\begin{figure*}
    \centering
     \includegraphics[width=0.8\textwidth]{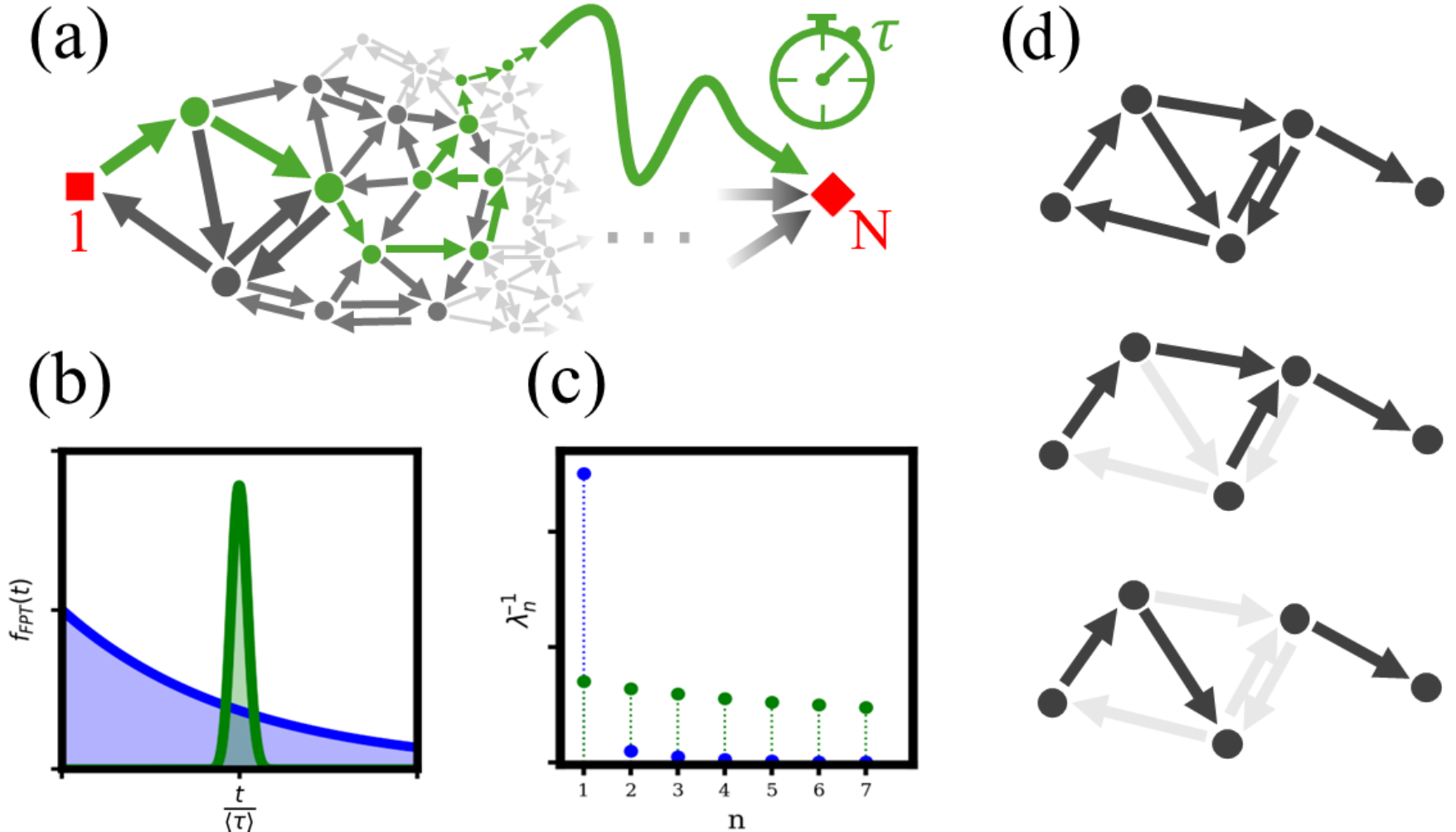}
    \caption{Generic first-passage time (FPT) distributions for large Markovian networks. (a) FPT in a large network: a trajectory starting at an initial state (square) reaches the target state $N$ (diamond) for the first time. (b) Two generic distributions have been observed before to emerge for kinetic proofreading
    networks in the limit of very large systems: the delta and the exponential distributions.   
    (c) Here we show that in the limit $N\to\infty$, the FPT-distribution is determined by the eigenvalue structure of the generator matrix.
    If infinitely many eigenvalues contribute, the distribution collapses to a delta peak. 
    If $\lambda_1^{-1}$ dominates the sum of inverse eigenvalues, the  FPT rescaled by its mean converges to an exponential law. (d) Examples of graph
    theoretical concepts used here: a graph, 
    a spanning tree and a two-tree spanning forest. A graph-theoretical decomposition provides an exact representation of the FPT-moments and the Laplace transform of the FPT-density.}
    \label{fig:main_idea}
\end{figure*}

Reaching a certain state in a stochastic system is called a first passage
and the corresponding time is called the first-passage time (FPT) (Fig.~\ref{fig:main_idea}(a)).
Like the underlying state variable, it is a stochastic variable
with a certain distribution. The statistical properties of first-passage 
times play a central role in physics \cite{hanggi1990reaction, hofmann2003mean,benichou2014first,friedman2017quantum,hasegawa2022thermodynamic,wang2024first,kewming2024first,prech2025optimal}, chemistry \cite{szabo1980first,woods2024analysis, rao2025exact,rao2025exact2}, biology \cite{bressloff2014stochastic,metzler2014first, chou2014first,iyer2016first, polizzi2017mean,frey2019stochastic,kaufmann_electrostatic_2020,bebon2022first}, and economics \cite{black1976valuing,leland1996optimal,perello2011scaling}, where they characterize noise-driven event timing, barrier-crossing dynamics, and rare transitions in complex stochastic systems \cite{Honerkamp,Redner,vanKampen,condamin2007first}. 

Previous work has established that FPT-statistics in confined and diffusion-controlled systems often exhibit robust and universal features. In particular, seminal work by Bénichou, Voituriez and collaborators \cite{benichou2010geometry,benichou2014first} showed that first-passage processes for diffusion in complex geometries can be classified in terms of transport properties, such as compact versus non-compact exploration, leading to universal scaling of mean FPTs and simple limiting distributions. Recent studies by Baravi, Kessler and Barkai \cite{baravi2025solutions,baravi2025first} showed that these FPT-distributions generically exhibit a characteristic biscaling structure with a short time dependence that reflects the initial condition and a universal long-time exponential tail.
This is in agreement with the work of Godec and Metzler, who identified the separation of time scales between direct and indirect trajectories as a mechanism leading to universal FPT-asymptotics \cite{godec2016first,godec2016universal}.

Here we address the question of universal limits of FPT-distributions for 
time-continuous and space-discrete master equations \cite{gillespie1992rigorous,murugan2012speed},
which are often used to describe networks of biochemical processes.
In the context of master equation models for kinetic proofreading,
it has been observed before, both through analytical and numerical
means, that on the timescale of the mean completion time,
the FPT-density tends to converge to one of two simple forms
in the limit of large network size: 
either to a delta distribution, indicating quasi-deterministic behavior, or 
to an exponential distribution, indicating quasi-memoryless 
behavior \cite{SimplicityComplexNW,munsky2009specificity} (Fig.~\ref{fig:main_idea}(b)).
As pointed out by the authors, this phenomenon helps to explain the success of coarse-grained Markov chain models for complex systems. It also raises the question how 
much microscopic information can be inferred from macroscopic observations.
In fact, it is common practice to infer system properties from experimental first-passage time measurements \cite{bomze2012noise, chung2018transition,thorneywork2020direct, broadwater2021first,zunke2022first, singh2025inferring,bayer2025real}. If only the delta distribution or the exponential distribution,
both of which are fully characterized by their means, emerged in the limit of
large networks, such inference methods would become impossible in the limit of large systems.
In the study of complex networks, ignorance about the microscopic details of the network remains a substantial obstacle to understanding these systems \cite{van2022thermodynamic,fritz2025entropy, maier2025observed}. Model independent results are therefore particularly desirable \cite{bebon2023controlling}. 

Interestingly, the two limiting distributions identified for kinetic 
proofreading networks also figure prominently in information theory and stochastic thermodynamics, 
because they correspond to the two extreme regimes of entropy for the first-passage time distribution: 
while the delta distribution is deterministic and contains no uncertainty, the exponential distribution can
be shown to correspond to the case of maximal entropy \cite{jarzynski1997equilibrium,jarzynski2012equalities,seifert2012stochastic, seifert2025stochastic}.
Thus both distributions make it difficult to infer the underlying microscopic dynamics,
but they must emerge for very different reasons.

Here we show how the two emerging limit distributions identified earlier for kinetic proofreading are in general related to the distribution of the eigenvalues of the generator matrix. We show that the key quantity is the sum of the inverse eigenvalues of the generator matrix. If infinitely many eigenvalues contribute comparably, a delta function emerges,
while the exponential limit results, when one value $\lambda_1^{-1}$ dominates (Fig.~\ref{fig:main_idea}(c)). \new{This can be related to a vanishing or a constant mean residual lifetime of the process compared to the mean first-passage time, respectively. We also identify more accessible criteria to identify the two limits, which moreover reveals that
they are asymmetric in nature.}

The deep connection between eigenspectra and properties of the FPT distribution - especially in the case where detailed balance holds for the steady state distribution, ensuring that the spectrum is real - has proven fruitful in the context of extreme value and time-average statistics \cite{hartich2019extreme, lapolla2020spectral}. 
Here we analyze this connection using the interpretation of first-passage times of Markovian networks in terms of graph theory, which recently gained more interest in the context of first-passage time problems \cite{NamPhD,nam2022linear, nam2023linear,khodabandehlou2022trees,khodabandehlou2023nernst,khodabandehlou2024vanishing,voits2025generic,haque2024graph,nam2025algebraic}. In particular, we will use the fact
that first-passage time moments can be obtained from a decomposition of the 
Markovian network into spanning trees and two-tree spanning forests (Fig.~\ref{fig:main_idea}(d)). We illustrate and validate our theoretical analysis by stochastic computer simulations using the Gillespie algorithm.

This paper is structured as follows. Sec.~\ref{sec:FPT_theory} gives an overview on FPT-theory for Markovian networks, including the graph theoretical interpretation and general results on the spectrum of the generator matrix. In Sec.~\ref{subsec:Macroscopic_forest_condition}, we consider the limit of large networks by relating the dependence of the cumulants of the FPT to the distribution of the eigenvalues depending on the limiting ratio of the weights of the two-tree spanning forests of the network. Then, we give sufficient conditions to conclude convergence to the deterministic and exponential limits in Secs.~\ref{subsec:Deterministic_limit} and \ref{subsec:Exponential_limit}, respectively. \new{A characterization based on the mean residual lifetime turns out to cover both regimes, while a simpler criterion based on comparing the ratios of the mean first-passage times in forward and backward direction is sufficient to guarantee convergence to the exponential limit, but fails to predict the deterministic regime.} Finally, we discuss this and other caveats to the universality of FPT distributions for large networks in Sec.~\ref{sec:Discussion}, uncovering which systems will lead to simple, large-scale statistics and which ones will not. Computer simulations with the Gillespie algorithm will be used throughout 
to illustrate and validate our theoretical results.

\section{First-Passage Time Theory for Markovian Networks}

\begin{figure*}[t]
    \centering
    \includegraphics[width=0.8\textwidth]{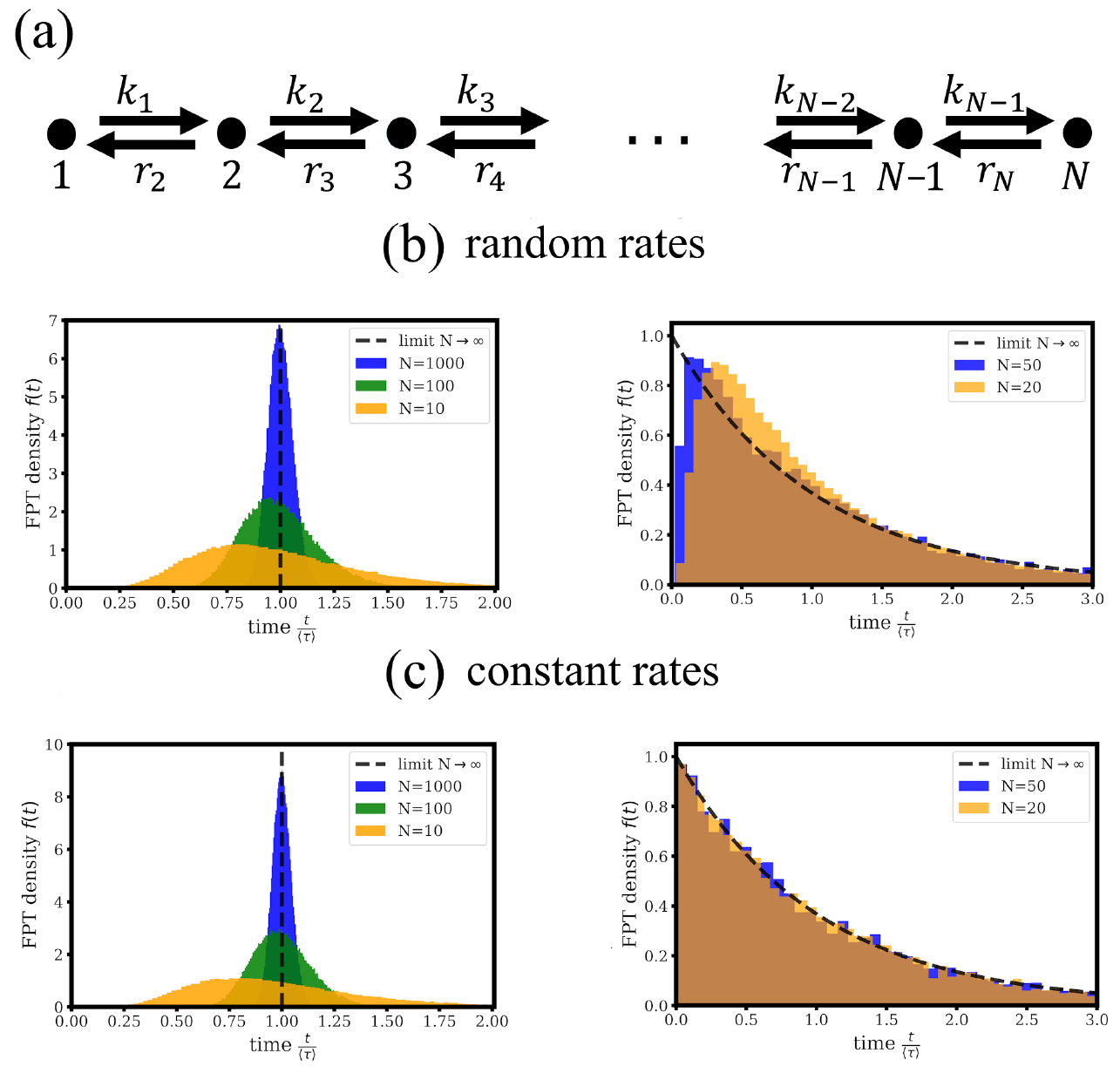}
    \caption{\new{The first-passage time (FPT) of a one-step master equation compared to its mean as system size $N$ increases. (a) The network of the one-step master equation with site-dependent forward rates $k_i$ and backward rates $r_i$. (b) FPT-statistics for $10^5$ runs with randomly drawn rates. Left: rates drawn from uniform distributions, $k_i\sim U[0.5,2]$, $r_i\sim U[0,1]$, showing convergence to a delta-distribution (deterministic limit). Right: rates drawn from uniform distributions, $k_i\sim U[0.5,1]$, $r_i\sim U[0.5,1.5]$, showing convergence to an exponential distribution (exponential limit). (c) FPT-statistics for $10^5$ runs with constant rates. Left: rates chosen as $k_i=\frac{3}{2}$, $r_i=\frac{1}{2}$, showing convergence to a delta-distribution (deterministic limit). Right: rates chosen as $k_i=\frac{3}{4}$, $r=1$, showing convergence to an exponential distribution (exponential limit).}}
    \label{fig:One_Step}
\end{figure*}

\label{sec:FPT_theory}
\subsection{Master equation}
\label{subsec:master_equation}
Consider a time-homogeneous jump process on $N$ states governed by the master equation  
\begin{equation}
\dot{p}_i(t) = \sum_{j=1}^{N} k_{ji}p_j(t) - k_{ij}p_i(t),\label{eq:master_equation}
\end{equation}
with $p_i(t)$ the probability of being in state $i$ and $k_{ij}$ the jump rate from $i$ to $j$. Defining  
\begin{equation}
\hat{K}_{ij} = 
\begin{cases}
k_{ji}, & i \neq j, \\
-\sum_m k_{im}, & i=j,
\end{cases}
\end{equation}
the equation becomes $\dot{\mathbf{p}}(t) = \hat{K}\mathbf{p}(t)$, with $\sum_{i=1}^{N} p_i(t) = 1$ and $\hat{K}$ singular. The discussion can be restricted to irreducible Markov chains, so that $\hat{K}$ is guaranteed to have a simple zero eigenvalue, and all other eigenvalues have a negative real part. The corresponding eigenvector is the steady-state distribution $\mathbf{\pi}$. One calls the process reversible, if $\mathbf{p}_{s}$ obeys the detailed balance condition, $k_{ij}\new{\pi_{i}}=k_{ji}\new{\pi_{j}}$ $\forall i,j$. Then $\hat{K}$ is similar to a symmetric matrix $\hat{K}^s=P\hat{K}P^{-1}$, where $P=\text{diag}(\sqrt{\new{\pi}_{1}},...,\sqrt{\new{\pi}_{N}})$, which is indeed symmetric, since for $i\neq j$:
\begin{align}
    \hat{K}^{s}_{ij}&=k_{ij}\sqrt{\frac{p_i}{p_j}}=k_{ji}\sqrt{\frac{p_j}{p_i}}=\hat{K}^s_{ji}.
\end{align} 
In particular, then all eigenvalues of $\hat{K}$ are real.

\subsection{Example: one-step master equation}
\label{subsec:Example_one_step_master_equation}

As an illustrative example, we consider a one-step master equation or birth-death process:
\begin{align}
    \dot{p}_i=-(k_i+r_i)p_i+k_{i-1}p_{i-1}+r_{i+1}p_{i+1},\label{eq:one_step_master_equation}
\end{align}
\new{for $0 < i < N$, and 
\begin{align}
        \dot{p}_0=-k_0p_0+r_{1}p_{1},\qquad\dot{p}_N=-r_Np_N+k_{N-1}p_{N-1},
\end{align}
i.e., a setting $i=0$ and $i=N$ as reflecting boundaries,}
and with finite rates ($k_\text{min}\leq k_i\leq k_\text{max}$, $r_\text{min}\leq r_i\leq r_\text{max}$ for all $i$). The steady state obeys the detailed balance condition, $k_{i}\new{\pi}_{i}=r_{i+1}\new{\pi}_{i+1}$ for all $i$ due to the reflecting boundary.
The corresponding network is shown in Fig.~\ref{fig:One_Step}(a). 

It is well known that the mean first-passage time (MFPT) to reach state 
$N$ starting from state $m$ can be expressed as
\begin{align}
    \langle \tau \rangle_{m \to N}= \sum_{i=m}^{N-1}\sum_{j=1}^{i}\frac{1}{k_j}\prod_{l=j+1}^{i}\frac{r_l}{k_l}\label{eq:MFPT_master_equation}.
\end{align}
This result is typically derived using recursive arguments \cite{vanKampen,gardiner2004handbook}, but can also be obtained directly within the graph-theoretical framework \cite{nam2022linear,voits2025generic}.
In contrast, determining the full FPT time distribution is substantially more challenging. Exact solutions exist only for particular choices for the rates \cite{redner2001guide,yin2012continuous}, while most available results rely on approximations \cite{li2013mechanisms,li2014pathway,smith2015general,assaf2017wkb,van2017orthogonal,kononovicius2019approximation}. 

As an alternative to analytical calculations, FPT-distributions can be efficiently sampled using stochastic simulations based on the Gillespie algorithm \cite{gillespie1977exact,gillespie2007stochastic}. 
In Fig.~\ref{fig:One_Step}(b), we show the results of such simulations 
for the one-step master equation from Eq.~\ref{eq:one_step_master_equation}.
We first use random rates drawn from uniform distributions $k_i\sim U[0.5,2]$, $r_i\sim U[0,1]$,
so that the \new{local rates are biased toward the target}. System size $N$ is increased from $10$
through $100$ to $1000$. One nicely sees how the
FPT distribution becomes sharply peaked, approaching a delta-like form 
on the scale of its mean. In marked contrast, if the rates are drawn
from $k_i\sim U[0.5,1]$, $r_i\sim U[0.5,1.5]$, \new{the local rates are biased away from the target} and the distribution converges toward an exponential form. 

Fig.~\ref{fig:One_Step}(c) shows the same FPT statistics for constant rates instead of randomly chosen ones, with $(k_i,r_i)=(\frac{3}{2}, \frac{1}{2})$ to get a forward bias and $(k_i,r_i)=(\frac{3}{4}, 1)$ to get a backward bias. The limiting behavior of the distributions is qualitatively identical to the previous case of random rates, showing that the limiting behaviors are not a consequence of randomly assigning the rates. This confirms earlier results for kinetic proofreading
networks that these two distributions tend to emerge in the limits of large system sizes, and that they can have a relation
to the bias in the system \cite{SimplicityComplexNW,munsky2009specificity}.

\subsection{First-passage time and graph theory}

\new{The first-passage time (FPT) $\tau_{i\to N}$ to state $N$ under the initial condition $p_{0,j}=\delta_{ij}$ describes the time it takes the process to reach $N$ for the first time. As the subsequent dynamics are irrelevant for the FPT, $N$ can be chosen as an absorbing state, i.e., a state without outgoing edges. In the master equation, this implies that there is no probability influx from state $N$, so} one can restrict \new{Eq.~(\ref{eq:master_equation})} to the first $N-1$ components:  
\begin{equation}
\dot{\mathbf{p}}(t) = K\mathbf{p}(t),
\end{equation}
where $K$ is the $N$-th principal minor of $\hat{K}$. \new{The solution for the first $N-1$ components is then given by:
\begin{align}
    \mathbf{p}(t)=e^{Kt}\mathbf{p}_0=e^{Kt}\hat{e}_i,
\end{align}
where $\hat{e}_i$ is the unit vector in $i$-direction, and the occupation probability for state $N$ then follows as $p_N(t)=1-\sum_{i=1}^{N-1}p_i(t)$. The process does not leave $N$ anymore once this state is reached since it was set to be an absorbing state. Therefore, the occupation probability now coincides with the distribution function of the FPT:
\begin{align}
    p_N(t)=p(\tau_{i\to N}\leq t),   
\end{align}
so its derivative yields the FPT density:
\begin{align}
    f_{i\to N}(t)&=\dot{p}_N(t)\\ &=-\sum_{i=1}^{N-1}\dot{p}_i(t)\\ &=-\mathbf{1}^T\dot{\mathbf{p}}(t)\\ &=-\mathbf{1}^TK e^{Kt}\hat{e}_{i}\\&=-\hat{e}_{i}e^{K^Tt}K^T\mathbf{1},
\end{align}
where the index $i$ indicates the conditioning on $p_{0,j}=\delta_{ij}$, the subscript $T$ denotes the transpose and $\mathbf{1}$ the vector with ones in all components. Defining $f_{0,i} :=-\sum_{j=1}^{N-1}K^T_{ij}= k_{iN}$, it follows that the} FPT density obeys the adjoint master equation \cite{Redner,vanKampen}:  
\new{\begin{equation}
\mathbf{f}(t) = e^{K^T t} \mathbf{f}_0 \quad \Rightarrow \quad \dot{\mathbf{f}}(t) = K^T \mathbf{f}(t),
\end{equation}
where the components of $\mathbf{f}(t)$ are $f_{i\to N}$.}
This implies that $\mathbf{f}(t)$ belongs to the class of phase-type distributions \cite{MatrixExpDistributions}. 

\new{Taking the Laplace transform $\mathcal{L}{\mathbf{f}}(s)$ of $\mathbf{f}(t)$ turns the differential equation into an algebraic relation:
\begin{align}
    s\mathcal{L}{\mathbf{f}}(s)- \mathbf{f}_0= K^T\mathcal{L}{\mathbf{f}}(s),             
\end{align}
which implies that:}  
\begin{equation}
\mathcal{L}{\mathbf{f}}(s) = (sI-K^T)^{-1} \mathbf{f}_0 = \sum_{n=0}^\infty s^n (K^{T})^{-n} \mathbf{1},
\end{equation}
so the $n$-th FPT moment is given by: 
\begin{equation}
\boldsymbol{\tau}^{(n)} =n(-K^{T})^{-1}\boldsymbol{\tau}^{(n-1)}= n! \, ((-K^T)^{-1})^n \mathbf{1}\label{eq:FPT_moments}.
\end{equation}

\begin{figure}[t]
    \centering
    \includegraphics[width=0.48\textwidth]{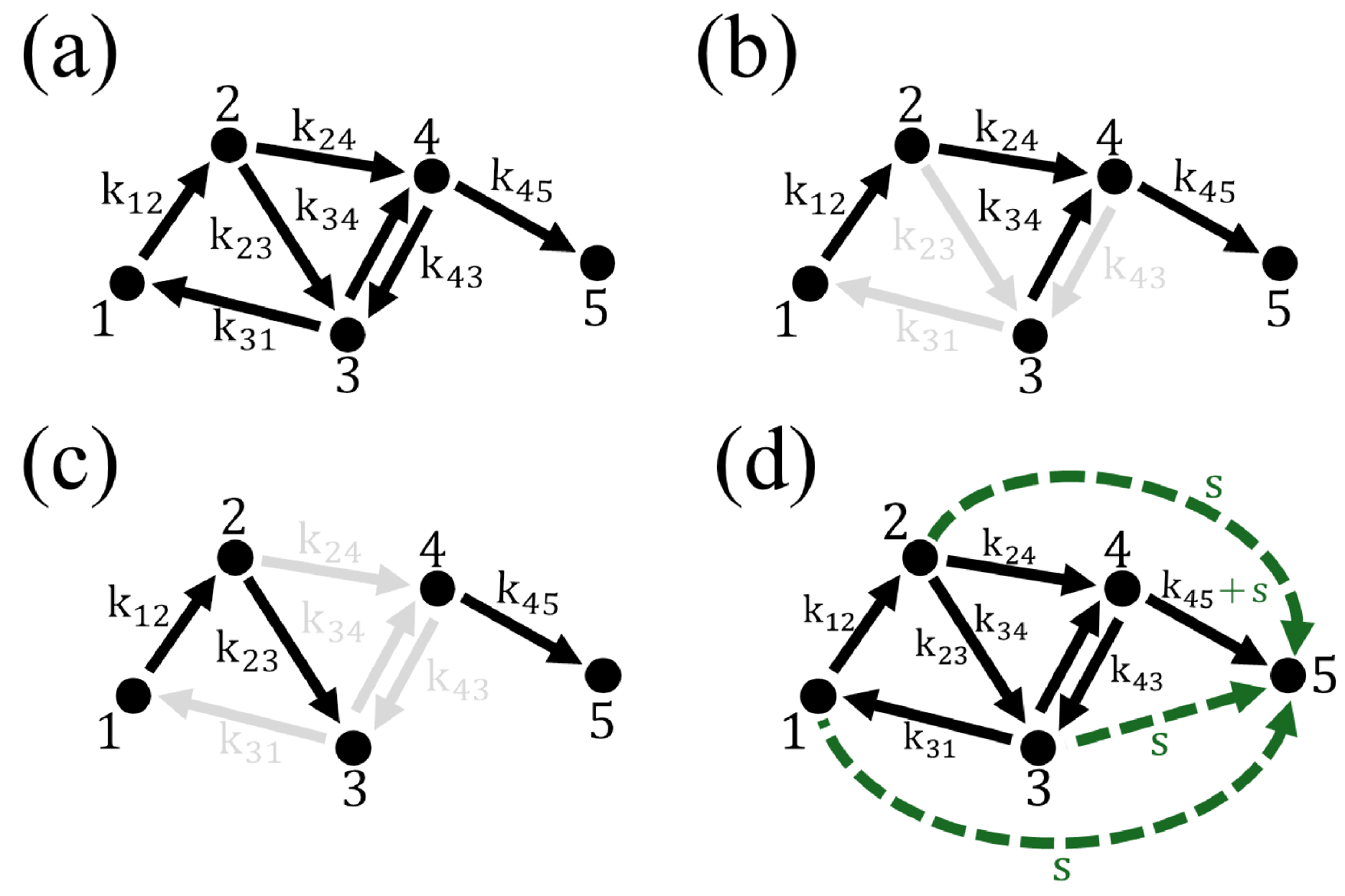}
    \caption{Illustration of graph theoretical concepts. (a) Example of a Markovian network. (b) A spanning tree of the graph with root in $5$. (c) A two-tree spanning forest with roots in $3$ and $5$. (d) Network with an outgoing edge to $5$ with weight $s$ added to all vertices, which can be used to calculate the Laplace transform of the first-passage time.}
    \label{fig:graph_tree_forest_augmented}
\end{figure}

\new{Note that the target state can always be set to be state $N$, without loss of generality. In the following, we will keep the target state implicit in the notation, with the understanding that we are always considering the first-passage time to $N$, i.e., we shall write $\tau_i$ and $f_{i}(t)$ instead of $\tau_{i\to N}$ and $f_{i\to N}(t)$.} For notational clarity, we write $\tau^{(1)}_i=\langle\tau_m\rangle$ for the mean first-passage time (MFPT) and define $M:=(-K^{T})^{-1}$ in the following. The components of this matrix have a graph-theoretical interpretation \new{if the Markovian network is interpreted as weighted and directed graph $G$, with the weights corresponding to the transition rates, which originates} from the All-Minors Matrix-Tree Theorem, a generalization of Kirchhoff's theorem (see also Supplement I) \cite{chaiken1982combinatorial,NamPhD,nam2023linear}:
\begin{align}
    M_{ij}=\frac{\sum_{\mathcal{F}^{i\to j}_{[j,N]}}w(\mathcal{F})}{\sum_{\mathcal{T}_{[N]}}w(\mathcal{T})}\label{eq:M_matrix}
\end{align}
where \new{$\sum_{\mathcal{T}_{[N]}}$} denotes the sum over all spanning trees rooted at $N$ and \new{$\sum_{\mathcal{F}^{i\to j}_{[j,N]}}$ is the sum over} the two-tree spanning forests with one tree rooted at $N$ and the other one rooted at $j$ containing $i$. The weight \new{$w(S)$ of a subgraph $S\subseteq G$ is defined as the product over all of its edges (rates) $E(S)\subseteq E(G)$, i.e., $w(S):=\prod_{k_{ij}\in E(S)}k_{ij}$}. In particular, this implies that $M_{ij}$ is non-negative $\forall i,j$ and all components are positive if and only if the \new{base graph $G$ from which the spanning trees and forests are taken} is strongly connected (ergodic) except for maybe $N$ (there is a path from any vertex excluding $N$ to any other vertex). Applying this identity to Eq. (\ref{eq:FPT_moments}), one  obtains, for instance, for the mean:
\begin{align}
    \langle\tau_m\rangle=\frac{\sum_{j=1}^{N-1}\sum_{\mathcal{F}^{m\to j}_{[j,N]}}w(\mathcal{F})}{\sum_{\mathcal{T}_{[N]}}w(\mathcal{T})}\label{eq:MFPT},
\end{align}
and one can also get similar expressions for higher moments.
In Fig.~\ref{fig:graph_tree_forest_augmented}(a), we show as an example 
a Markovian network with five vertices. Fig.~\ref{fig:graph_tree_forest_augmented}(b) and (c) show examples for a spanning tree and for a two-tree spanning forest of that network, respectively.

An immediate corollary \new{of} the graph-theoretic interpretation is that $M_{ij}\leq M_{jj}$ $\forall i,j$, since every tree rooted at $j$ containing $i$ is also a tree rooted at $j$ containing $j$. \new{This relation can be used to provide an upper bound for the mean first-passage time in terms of the eigenvalues $\lambda_1,...,\lambda_{N-1}$ of $-K^T$, labeled such that $\text{Re}(\lambda_1)\leq\text{Re}(\lambda_2)\leq...\leq\text{Re}(\lambda_{N-1})$. Eq.~(\ref{eq:FPT_moments}) for $n=1$ implies:}
\begin{align}
    \max_i\langle\tau\rangle_i&=\max_i\Big(\sum_jM_{ij}\Big)\leq\new{\sum_jM_{jj}=}\text{tr}(M)=\new{\sum_{i=1}^{N-1}}\frac{1}{\lambda_i},\label{eq:upper_bound_max_tau}
\end{align}
\new{using that the trace of $M=(-K^T)^{-1}$ is the sum of the inverse eigenvalues of $-K^T$.}
We next observe that $s\mathds{1} -K^T$ is of the same shape as $-K^T$ if one replaces \new{$k_{iN}\to k_{iN} + s$}, as illustrated in Fig.~\ref{fig:graph_tree_forest_augmented}(d). So $s\mathds{1} -K^T$ can be inverted by the same logic:
\begin{align}
    ( s\mathds{1} -K^T)^{-1}_{ij}=\frac{\sum _{\mathcal{F}_{s,[j,N]}^{i\rightarrow j}}w(\mathcal{F})}{\sum_{\mathcal{T}_{s,[N]}}w(\mathcal{T})}\label{eq:forsets_trees_augmented_graph}.
\end{align}
The index $s$ indicates that here the $s$ rates have to be included. In particular, note that the denominator is the characteristic polynomial of $K^T$:
\begin{align}
    \sum_{\mathcal{T}_{s,[N]}}w(\mathcal{T})=\det(s\mathds{1} -K^T)=\prod_{i=1}^{N-1}(s+\lambda_i).\label{eq:CharPolynomial}
\end{align}

For the FPT density, one then finds:
\begin{align}
    \mathcal{L}{f}_m(s) &= \frac{\sum_{j=1}^{N-1} k_{j,N}\sum _{\mathcal{F}_{s,[j,N]}^{m\rightarrow j}}w(\mathcal{F})}{\sum_{\mathcal{T}_{s,[N]}}w(\mathcal{T})}\label{eq:Laplace_Transform_FPT_Graph_Theory}\\ &= \frac{\sum_{j=1}^{N-1} k_{j,N}\sum _{\mathcal{F}_{s,[j,N]}^{m\rightarrow j}}w(\mathcal{F})}{\sum_{n=0}^{N-1}\Big(\sum _{\substack{|\mathcal{F}_{[N]}| =N-1-n}}w(\mathcal{F})\Big)s^n},\label{eq:Laplace_FPT_density}
\end{align}
where the second sum in the denominator denotes the sum over all spanning forests with one root in $N$ and $N-1-n$ edges (meaning that it is an $n$-tree forest). \new{This identity follows from observing that every spanning forest of the base network, where $N$ is one of the roots, turns into a spanning tree in the augmented network by connecting all other roots to $N$ with an $s$-rate, and conversely, every such spanning tree of the augmented network turns into a spanning forest of the base network if the $s$-rates are removed.}

\section{Limit of large Networks}
\label{sec:Limit_of_large_networks}

\subsection{Perron-Frobenius theorem and an intuitive argument}

The Perron-Frobenius theorem guarantees the existence of a simple, dominant real eigenvalue, which is known as the Perron-Frobenius eigenvalue, for positive and, more generally, for non-negative matrices (with strict dominance in the positive case) \cite{meyer2023matrix}. 
This theorem is a cornerstone in the study of Markov chains, as the existence of a steady-state solution, and its uniqueness for irreducible chains, follow directly from it \cite{seneta2006non,van1992stochastic}. 

As argued above, $M$ is positive if the network is strongly connected, so $\frac{1}{\lambda_1}$ corresponds to the Perron--Frobenius eigenvalue. 
Furthermore, it is bounded by the smallest and largest row sums of $M$ \cite{deutsch1981bounds}:
\begin{align}
    \min_i \langle \tau_i \rangle \leq \frac{1}{\lambda_1} \leq \max_i \langle \tau_i \rangle.
\end{align}
Combining this with Eq. (\ref{eq:upper_bound_max_tau}) yields upper and lower bounds for the maximum MFPT:
\begin{align}
   \frac{1}{\lambda_1} \leq \max_i \langle \tau_i \rangle \leq \sum_i \frac{1}{\lambda_i}.
   \label{eq:MFPT_bounds}
\end{align}
\new{For large networks, we are interested in the behavior of the eigenvalues in the limit $N \to \infty$. Consider first the case that the eigenvalue $\lambda_1$ rapidly goes to zero, while the other eigenvalues stay finite or converge much slower to zero. Mathematically, this means that:}
\begin{align}
    \lim_{N \to \infty} 
    \frac{\frac{1}{\lambda_1}}{\sum_i \frac{1}{\lambda_i}} = 1.
\end{align}
Then the MFPT scales as $\langle \tau \rangle := \max_i \langle \tau_i \rangle \sim \frac{1}{\lambda_1}$, and one expects $\frac{1}{\langle \tau \rangle} M \sim P$, where $P$ is the projection operator onto the Perron-Frobenius eigenspace. 
Using Eq. (\ref{eq:FPT_moments}), this implies for the moments:
\begin{align}
    \boldsymbol{\tau}^{(n)} 
    \sim n!  P^n \mathbf{1} 
    = n!  P \mathbf{1} 
    = n!  \boldsymbol{\tau}^{(1)},
\end{align}
since the projection operator satisfies $P^2 = P$. 
Thus, one obtains exactly the moments of an exponential distribution.

Conversely, if $\frac{1}{\lambda_1} \ll \langle \tau \rangle$, one expects 
\begin{align}
    \langle \tau \rangle \sim \sum_i \frac{1}{\lambda_i},
    \qquad 
    \sigma_\tau^2 = \langle \tau^2 \rangle - \langle \tau \rangle^2 
    \sim \sum_i \frac{1}{\lambda_i^2},
\end{align}
which leads to:
\begin{align}
    \frac{\sigma_\tau^2}{\langle \tau \rangle^2} 
    \sim 
    \frac{\sum_i \frac{1}{\lambda_i^2}}
         {\left( \sum_i \frac{1}{\lambda_i} \right)^2}
    \overset{N \to \infty}{\longrightarrow} 0,
\end{align}
provided that infinitely many terms $\frac{1}{\lambda_i}$ contribute non-negligibly to the sum. 
In this case, the first-passage process approaches a deterministic limit as $N \to \infty$.

While this reasoning is heuristic - for instance, it ignores possible changes in the Perron-Frobenius eigenvector as $N \to \infty$ - it nevertheless provides valuable intuition for the analysis that follows, where we establish under which conditions this argument is valid and discuss cases in which it fails.

\subsection{Macroscopic forest condition}
\label{subsec:Macroscopic_forest_condition}

Intuitively, the large-network limit can yield meaningful macroscopic behavior only if the initial state $m$ is sufficiently far from the target state $N$. 
If the system starts too close to $N$, a small, local subnetwork may dominate the first-passage dynamics, leading to non-universal asymptotic behavior characteristics of the full system. 
As argued in the previous section, the deterministic and exponential limits depend on the distribution of the eigenvalues. 
We now translate this insight into a mathematically precise condition. 

First, observe that
\begin{align}
    \frac{1}{1+\frac{\sum_{j=1}^{N-1}\sum_{\mathcal{F}^{m\to j}_{[j,N]}}w(\mathcal{F})}{\sum_{j=1}^{N-1}\sum_{\mathcal{F}^{m\to N}_{[j,N]}}w(\mathcal{F})}}=1-\frac{\langle\tau_m\rangle}{\sum_{i=1}^{N-1}\frac{1}{\lambda_i}},
\end{align}
\new{where $\mathcal{F}^{m\to N}_{[j,N]}$ are the spanning forests of two trees with roots $j$ and $N$ where $m$ is on the tree rooted at $N$. This} follows from Eq. (\ref{eq:MFPT}) and from the identity 
\new{\begin{align}
    \mathrm{tr}(M) &= \sum_{j=1}^{N-1}\sum_{\mathcal{F}_{[j,N]}} w(\mathcal{F})\\ & = \sum_{j=1}^{N-1}\big(\sum_{\mathcal{F}^{m\to j}_{[j,N]}} w(\mathcal{F})+\sum_{\mathcal{F}^{m\to N}_{[j,N]}} w(\mathcal{F})\big),
\end{align}
where $\mathcal{F}_{[j,N]}$ are all the two-tree spanning forests with roots $j$ and $N$, which can always be split into the spanning forests where $m$ belongs to the $j$-tree and where $m$ belongs to the $N$-tree.
}
Thus, one concludes that
\begin{align}
    &r:=\lim_{N\to\infty}\frac{\sum_{j=1}^{N-1}\sum_{\mathcal{F}^{m\to N}_{[j,N]}}w(\mathcal{F})}{\sum_{j=1}^{N-1}\sum_{\mathcal{F}^{m\to j}_{[j,N]}}w(\mathcal{F})}<\infty \\\Leftrightarrow&\lim_{N\to\infty}\frac{\langle\tau_m\rangle}{\sum_{i=1}^{N-1}\frac{1}{\lambda_i}}=\frac{1}{1+r}>0.
\end{align}

Therefore, if the ratio $r$ of the two-tree spanning forests in which the vertex $m$ lies on the $N$-tree to those in which it lies on the complementary tree remains finite as $N \to \infty$, 
then the mean first-passage time (MFPT) from $m$ to $N$ scales with the sum of the eigenvalue inverses of $M$. 
Note that this condition reflects the requirement that $m$ must not be located too close to $N$, since otherwise $m$ would almost always belong to the $N$-tree, violating the macroscopic scaling assumption.

An upper bound on $r$ with a more intuitive interpretation is given by
\begin{align}
    r \leq \max_{j=1,\ldots,N-1} \frac{p_{m\to\{j,N\}}(N)}{1-p_{m\to\{j,N\}}(N)},
\end{align}
where $p_{m\to\{j,N\}}(N)$ denotes the splitting probability that a process starting at $m$ reaches $N$ before reaching the other vertex $j$. This bound implies that $r$ remains finite in the limit $N \to \infty$, unless there exists another vertex whose probability of being reached before vertex $N$ vanishes in that limit.

Using Eq. (\ref{eq:Laplace_Transform_FPT_Graph_Theory}) to express the Laplace transform of the FPT density with time normalized to the mean, one finds for the logarithm of $\mathcal{L}f(s)$:
\begin{align}
    \ln(\mathcal{L}{f}(s))=\sum_{n=1}^\infty\frac{\sum_{i=1}^{N-1}\frac{1}{\lambda_i^n}}{\langle\tau_m\rangle^n}\frac{(-s)^n}{n}+\ln\sum_{n=0}^{N-1}r_ns^n,
\end{align}
where:
\begin{align}
    r_n:=\frac{\sum_{i_1,...,i_n} \sum _{\mathcal{F}_{[i_1,...,i_n,N]}^{m\rightarrow N}}w(\mathcal{F})}{\langle\tau_m\rangle^n\sum_{\mathcal{T}_{[N]}}w(\mathcal{T})}.
\end{align}
Observe that Eq. (\ref{eq:MFPT}) implies that $r_1=r$. Exploiting the graph theoretical interpretation of the coefficients, one can show that for any $N$ holds (see Supplement II):
\begin{align}
    1\leq&\sum_{n=0}^{N-1}r_ns^n \leq \frac{1-s}{1-(1+r)s},
\end{align} 
for $s<\frac{1}{1+r}$. In particular, if $r=0$, this implies:
\begin{align}
    \ln(\mathcal{L}{f}(s))=\sum_{n=1}^\infty\lim_{N\to\infty}\frac{\sum_{i=1}^{N-1}\frac{1}{\lambda_i^n}}{\langle\tau_m\rangle^n}\frac{(-s)^n}{n},
\end{align}
and one concludes that the $n$-th cumulant is given by:
\begin{align}
    \lim_{N\to\infty}\kappa_n \Big(\frac{t}{\langle\tau_m\rangle}\Big) =(n-1)!\lim_{N\to\infty}\frac{\sum_{i=1}^{N-1}\frac{1}{\lambda_i^n}}{\Big(\sum_{i=1}^{N-1}\frac{1}{\lambda_i}\Big)^n}.
\end{align}

This result is the key insight that allows us to understand the two limiting behaviors: if there exists a single dominant eigenvalue of $-K^{-1}$ in the limit $N\to\infty$ such that $\frac{\frac{1}{\lambda_1}}{\sum_i\frac{1}{\lambda_i}}\to 1$, then the cumulants become $\kappa_n\to(n-1)!$, so the FPT distribution converges to an exponential distribution. Conversely, if infinitely many eigenvalues remain of comparable magnitude, the cumulants become $\kappa_n\to\delta_{1n}$. So the distribution collapses to a delta peak, corresponding to the deterministic limit. 
In the following, we analyze these two regimes separately and derive explicit conditions on the eigenvalue spectrum under which each limiting behavior arises.

Note that if there is a unique path from $m$ to $N$ that contains all  vertices (i.e., it is a Hamiltonian path), then there cannot be any spanning forest of more than one tree with a $m\to N$ path. Hence, the macroscopic network condition is true and yields $r=0$ even before taking the limit $N\to\infty$. This is the case in the example of the one-step master equation from Sec. \ref{subsec:Example_one_step_master_equation} when choosing $m=1$ and also for the kinetic proofreading networks \cite{SimplicityComplexNW,munsky2009specificity}.

\subsection{Deterministic limit}
\label{subsec:Deterministic_limit}
For the deterministic limit, it is necessary and sufficient that the variance $\kappa_2 \big(\frac{t}{\langle\tau_m\rangle}\big)=\sigma^2_m$ vanishes as $N\to\infty$. If the macroscopic forest condition holds with $r=0$, this requires:
\begin{align}
    \frac{\sum_{i=1}^N\frac{1}{\lambda_i^2}}{\Big(\sum_{i=1}^N\frac{1}{\lambda_i}\Big)^2}\overset{N\to\infty}{\longrightarrow}0.
\end{align} 
In Supplement III, we prove that this holds if $\langle\tau_m\rangle\overset{N\to\infty}{\longrightarrow}\infty$ and $\lim_{N\to\infty}\lambda_1\geq M>0$. The former condition that the MFPT from $m$ to $N$ diverge simply reflects the fact again that one has to start macroscopically far away from the target $N$ in order for the large network limit to give any generic results. The latter condition requires that the smallest eigenvalue $\lambda_1$, which is real as argued before, stay bounded away from $0$ as $N\to\infty$. 

If the network is strongly connected, that is, any two vertices are connected by a path, one can use a result by Lawler and Sokal \cite{lawler1988bounds} who generalized a famous result by Cheeger \cite{cheeger1970lower}. If the graph of $\hat{K}$ is strongly connected, then holds for the eigenvalue of $\hat{K}$ with the second smallest real part $\mu_2$: 
\begin{align}
     \frac{\Phi^2}{8c}\leq\text{Re}(\mu_2),
\end{align}
where $c:=\max_i\sum_jk_{ij}$ is the maximal sum of the outgoing rates from a vertex and the conductance $\Phi$ is defined as:
\begin{align}
    \Phi:=\min_{0<\new{\pi}(A)<1}\frac{\sum_{i\in A, j\in A^c}k_{ij}\new{\pi}_{i}}{\new{\pi}(A)\new{\pi}(A^c)},
\end{align}
where $\new{\pi}_{i}$ is the steady state distribution of the network and the minimum is taken over all partitions of the vertex set into $A$ and $A^c$. Note that, if the detailed balance condition holds, $A$ is a set minimizing $\Phi$ if and only if $A^c$ is a set minimizing $\Phi$.

\new{It is important to note that this is a statement on the eigenvalues of the unreduced generator matrix $\hat{K}$. However, relevant for our case are the eigenvalues of the $N$-th principal minor $K$, which only coincide with the non-zero eigenvalues of $\hat{K}$ if $N$ is
chosen to be absorbing, meaning that the network cannot be strongly connected anymore. To reconcile this, one can introduce a small back-rate $\epsilon$ from N to any other vertex $n$:
\begin{align}
    \hat{K}_{ij}\to \hat{K}_{ij}+\epsilon\delta_{in}\delta_{jN}-\epsilon\delta_{iN}\delta_{jN}
\end{align}
and let $\epsilon\to 0$. Note that taking the $N$-th principal minor always results in the same $K$ for any choice of $n$ and $\epsilon$ as both cannot change the FPT behavior to reach $N$.} 

We can apply this result to the one-step master equation from Sec. \ref{subsec:Example_one_step_master_equation} starting at $m=1$. Here, one sees that a minimizing set has to be of the shape $A=\{1,...,n\}$, which yields:
\begin{align}
    \frac{\sum_{i\in A, j\in A^c}k_{ij}\new{\pi}_{i}}{\new{\pi}(A)\new{\pi}(A^c)}\geq\frac{1}{\sum_{i=1}^n\frac{1}{k_i}\prod_{j=i+1}^n\frac{r_i}{k_i}}=\langle\tau\rangle_{n-1\to n}^{-1}
\end{align}
So, if the average time to advance one step is always finite, one can conclude convergence to the deterministic limit, explaining the convergence observed in Fig.~(\ref{fig:One_Step})(b).
\new{Note that a naive extension of this criterion to the global ratio of the MFPTs} is not sufficient, as illustrated in Sec. \ref{subsec:Forward_not_sufficient}. However, the intuition that stochastic dynamics biased to the target can be turned into a mathematically rigorous condition by considering the mean residual lifetime (MRL) of the process \cite{guess198812,hall2020estimation}:
\begin{align}
    \langle\tau\rangle(t)=\frac{\sum_{i=1}^{N-1}\langle\tau_i\rangle p_i(t)}{1-p_N(t)}, 
\end{align}
which describes the MFPT over time given that the process has not reached state $N$ yet. As a direct consequence of Eq.~(\ref{eq:FPT_moments}), one sees that the MRL obeys the following differential equation:
\begin{align}
    -1=\frac{\text{d}}{\text{d}t}\langle\tau\rangle(t)- \frac{\dot{p}_N(t)}{1-p_N(t)}\langle\tau\rangle(t).
\end{align}
The stationary solution for $t\to\infty$ is therefore given by:
\begin{align}
    \lim_{t\to\infty}\langle\tau\rangle(t)&= \lim_{t\to\infty}\frac{1-p_N(t)}{\dot{p}_N(t)}= -\lim_{t\to\infty}\frac{f(t)}{\dot{f}(t)}=\frac{1}{\lambda_1}\label{eq:MRL_limit},
\end{align}
using l'Hôpital's rule and the fact that $f(t)=\dot{p}_N(t)$ for choosing vertex $N$ as absorbing. The last equality is a result of the asymptotic scaling of the first-passage time density as $f(t)\sim e^{-\lambda_1t}$ \cite{godec2016universal}. Hence, if the MRL becomes small for long times relative to the MFPT (which is the MRL at $t=0$), this allows to conclude convergence to the deterministic limit. 
In the case that the network is not strongly connected, the FPT-distributions for the connected components have to be computed. The Laplace transform of the global FPT can then be expressed as:
\begin{align}
    \mathcal{L}{f}(s)=\sum_{i}P_i\prod_{j=1}^{n_i}\mathcal{L}{g}_{ij}(s),
\end{align}
where $P_i$ is the probability to take a specific path through the connected components and $\mathcal{L}{g}_{ij}(s)$ are FPTs of the connected components, hence satisfying $\mathcal{L}{g}_{ij}(0)=1$, $\frac{\text{d}\mathcal{L}{g}_{ij}}{\text{d}s}\Big|_{s=0}=-\langle\tau_{ij}\rangle$. Then $\langle\tau\rangle_i=\sum_{j}\langle\tau\rangle_{ij}$ is the MFPT along the $i$-th path so that $\langle\tau\rangle=\sum_iP_i\langle\tau\rangle_i$. If $\frac{\langle\tau\rangle_{ij}}{\langle\tau\rangle}\to0$ $\forall j$ as $N\to\infty$, one gets for $s\to\frac{s}{\langle\tau\rangle}$:
\begin{align}
    &\ln\Big(\prod_{j=1}^{n_i}\mathcal{L}{g}_{ij}\Big(\frac{s}{\langle\tau\rangle}\Big)\Big)=-\frac{\langle\tau\rangle_{i}}{\langle\tau\rangle}s+\mathcal{O}\Big(\Big(\frac{s}{\langle\tau\rangle}\Big)^2\Big),
\end{align}
so, as $N\to\infty$, one gets:
\begin{align}
    &\mathcal{L}{f}(s)=\sum_{i}P_ie^{-\frac{\langle\tau\rangle_{i}}{\sum_iP_i\langle\tau\rangle_i}s} \\\Rightarrow &f(t)=\sum_{i}P_i\delta\big(t{-\frac{\langle\tau\rangle_{i}}{\sum_iP_i\langle\tau\rangle_i}}\big),\label{eq:irreversible_limit_deltas}
\end{align}
so the FPT-density becomes a sum over delta functions weighted by the probabilities for the corresponding paths.

\begin{figure}[t]
    \centering
    \includegraphics[width=0.4\textwidth]{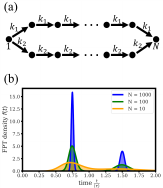}
    \caption{A minimal example to illustrate how irreversible transitions can result in two delta peaks. (a) A network branching into two Poisson processes with rates $k_1$ and $k_2$, both starting at vertex $1$ and ending in vertex $N$. (b) The first-passage time density results in two distinct delta peaks as $N$ increases, corresponding to the two chains. Here shown for $k_1=1$ and $k_2=2$.}
    \label{fig:two_peaks}
\end{figure}

As a minimal example, consider the network shown in Fig.~(\ref{fig:two_peaks}). It consists of two Poisson processes with rates $k_1$ and $k_2$, both going from vertex $1$ to vertex $N$. Since all transitions are irreversible, every vertex on its own is a connected component, so the global MFPT clearly dominates the individual steps. There are two possible paths with probabilities $P_1=\frac{k_1}{k_1+k_2}$ and $P_2=\frac{k_2}{k_1+k_2}$ with individual MFPTs $\langle\tau\rangle_1=\frac{N-1}{2k_1}$ and $\langle\tau\rangle_2=\frac{N-1}{2k_2}$, respectively. Thus, the MFPT is $\langle\tau\rangle=\frac{N-1}{k_1+k_2}$, and according to Eq. (\ref{eq:irreversible_limit_deltas}), the FPT density in the limit of $N\to\infty$ should read:
\begin{align}
    f(t)&=\frac{k_1}{k_1+k_2}\delta\Big(t-\frac{k_1+k_2}{2k_1}\Big)+\frac{k_2}{k_1+k_2}\delta\Big(t-\frac{k_1+k_2}{2k_2}\Big).
\end{align}
This result agrees well with intuition and also with the limit of the simulated first-passage time distributions shown in Fig.~\ref{fig:two_peaks}(b), which in this case is simply a weighted sum of two Erlang distributions.

\subsection{Exponential limit}
\label{subsec:Exponential_limit}

To establish convergence to the exponential limit, one must show that
\begin{align}
    \frac{\frac{1}{\lambda_1}}{\sum_{i=1}^{N-1}\frac{1}{\lambda_i}}
    \overset{N\to\infty}{\longrightarrow} 1.
\end{align}
Once this condition is met, the bounds on $\max_i\langle\tau_i\rangle$ (Eq. (\ref{eq:MFPT_bounds})) imply that
\begin{align}
    \frac{\max_i\langle\tau_i\rangle}{\sum_{i=1}^{N-1}\frac{1}{\lambda_i}}
    \overset{N\to\infty}{\longrightarrow} 1.
\end{align}
Thus, the macroscopic forest condition automatically holds with $r=0$ for 
$\langle\tau_m\rangle=\max_i\langle\tau_i\rangle$. 
However, proving the existence of such a dominant eigenvalue is generally nontrivial. 
In the following, we introduce two possible tests for this limit.

The first approach is to verify that the macroscopic forest condition holds with $r=0$. 
Unlike in the deterministic limit, convergence to an exponential distribution requires that 
all cumulants approach those of the exponential distribution for all orders $n$. 
The requirement that the coefficient of variation converges to unity,
\begin{align}
    \lim_{N\to\infty}\frac{\sigma_m^2}{\langle\tau_m\rangle^2}=1,
\end{align}
is therefore necessary but not sufficient.  
Although a coefficient of variation equal to $1$ is indicative of exponential form, it could also arise in other distributions.

However, if the macroscopic forest condition holds with $r=0$, this criterion becomes sufficient, as shown in Supplement IV. 
This result also justifies earlier work
on kinetic proofreading networks \cite{SimplicityComplexNW,munsky2009specificity}, which inferred exponential convergence from observing that the coefficient of variation tends to $1$.

A second test is based on the physical expectation proposed earlier that the first-passage time distribution approaches a delta or exponential form depending on whether the dynamics is \new{globally} biased toward or away from the target state, respectively \cite{SimplicityComplexNW,munsky2009specificity}.  
If the steady-state distribution satisfies detailed balance, \new{this can be turned into a simple criterion on the ratio of the MFPT in forward and backward direction. Namely}, for $\langle\tau\rangle_{m\to N}:=\max_{i}\langle\tau\rangle_{i\to N}$, one has
\begin{align}
    \frac{\langle\tau\rangle_{N\to m}}{\langle\tau\rangle_{m\to N}}
    \overset{N\to\infty}{\longrightarrow} 0
    \Rightarrow
    \lim_{N\to\infty}\kappa_{n}\!\left(\frac{t}{\langle\tau_m\rangle}\right)
    =(n-1)!,
\end{align}
provided that the macroscopic forest condition holds for the network 
with $m$ as the target and $N$ as the initial state with any finite $r$:
\begin{align}
    \lim_{N\to\infty}\frac{\sum_{j=1}^{N-1}\sum_{\mathcal{F}^{N\to m}_{[j,m]}}w(\mathcal{F})}{\sum_{j=1, j\neq m}^{N}\sum_{\mathcal{F}^{N\to j}_{[j,m]}}w(\mathcal{F})}=r< \infty.
\end{align}
In this case, the first-passage time distribution for $m\to N$ converges to an exponential form.
This follows from the interlacing property of the eigenvalues of different principal minors of $M$ when detailed balance holds \cite{hartich2019interlacing}. 
A formal proof is provided in the Supplement IV, where we exploit that $K$ is similar to a symmetric matrix via the diagonal similarity transformation $P=\text{diag}(\sqrt{\new{\pi}_{1}},\ldots,\sqrt{\new{\pi}_{N}})$, as explained above Sec. (\ref{subsec:master_equation}). The argument closely follows Fisk’s short proof of Cauchy’s interlacing theorem \cite{fisk2005very}.

This result applies directly to the one-step master equation. As explained in Sec. \ref{subsec:Example_one_step_master_equation}, its steady-state probability distribution satisfies the detailed balance condition. Moreover, the reverse transition $N\!\to\!1$ forms a Hamiltonian path, so the macroscopic forest condition also holds even with $r=0$. This rigorously establishes the exponential limit observed in Fig.~\ref{fig:One_Step}(c).

\subsection{\new{Naive global forward bias in terms of MFPT ratios is not sufficient for deterministic limit}}
\label{subsec:Forward_not_sufficient}

\begin{figure}[t]
    \centering
    \includegraphics[width=0.8\linewidth]{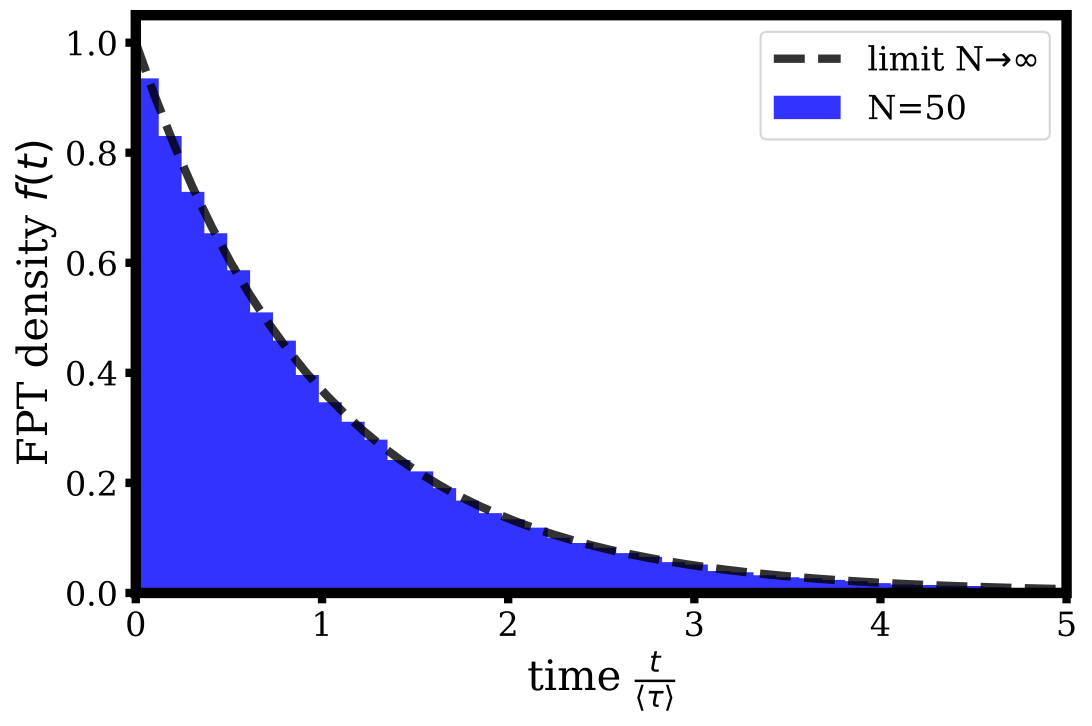}
    \caption{FPT-statistics for a one-step master equation with $N=50$ for $10^5$ trials with rates as specified in Eq. (\ref{eq:apparent_forward_bias}). \new{Although the ratio of the MFPTs suggests a naive global} forward bias, the limiting behavior of the distribution is exponential, showing that \new{a simple inversion of the previously identified condition for the exponential limit is not enough to }lead to a quasi-deterministic behavior.}
    \label{fig:apparent_forward_bias}
\end{figure}

Motivated by the result in the exponential regime, one might expect that the condition $ \frac{\langle\tau\rangle_{1\to N}}{\langle\tau\rangle_{N\to 1}}\overset{N\to\infty}{\longrightarrow}0$ could ensure deterministic behavior. The following counterexample demonstrates that this is not the case: consider the one-step master equation from Eq. (\ref{eq:one_step_master_equation}) with the rates
\begin{align}
    k_i=\begin{cases}
        3&\text{ if }i<\frac{N}{2}\\1&\text{ if }i\geq\frac{N}{2}
    \end{cases}
    , \hspace{1cm}
    r_i=\begin{cases}
        1&\text{ if }i<\frac{N}{2}\\\frac{3}{2}&\text{ if }i\geq\frac{N}{2}.
    \end{cases}\label{eq:apparent_forward_bias}
\end{align}
Inserting these into Eq. (\ref{eq:MFPT_master_equation}), one has:
\begin{align}
    \langle\tau\rangle_{1\to N}&=\Big(\frac{3}{2}\Big) ^{\frac{N}{2}+1}+\mathcal{O}(N)\\\langle\tau\rangle_{N\to 1}&=3 ^{\frac{N}{2}+1}+\mathcal{O}(N).
\end{align}
Therefore, $\lim_{N\to \infty}\frac{\langle\tau\rangle_{1\to N}}{\langle\tau\rangle_{N\to 1}}=0$. \new{While the criterion based on the ratio of the MFPTs would naively suggest a forward bias}, the FPT from state $1$ to state $N$ is dominated by the segment $\frac{N}{2}\to N$, where the system experiences a backward bias. This segment alone leads to an exponential limit, preventing convergence to the deterministic limit (see Fig.~\ref{fig:apparent_forward_bias}).

\new{Note that} the harmonic mean of the local ratios $\frac{k_i}{r_i}$ for this example reads:
\begin{align}
    \frac{N-2}{\sum_{i=2}^{N-1}\frac{r_i}{k_i}}=\frac{N-2}{\frac{N-2}{2}(\frac{1}{3}+\frac{3}{2})}=\frac{12}{11}>1.
\end{align}
Since the harmonic mean is always less than or equal to the geometric mean, which in turn is less than or equal to the arithmetic mean, we also have:
\begin{align}
    1<\Big(\prod_{i=2}^{N-1}\frac{k_i}{r_i}\Big)^\frac{1}{N-2}\leq\frac{1}{N-2}\sum_{i=2}^{N-1}\frac{k_i}{r_i}.
\end{align}
Thus, defining a 'forward bias' through these possible means of the ratios $\frac{k_i}{r_i}$ still fails to guarantee convergence to the deterministic limit \new{for the one-step master equation. In fact, it turns out that in this case, the correct classification is the logarithmic bias $\langle\log(k_i / r_i)\rangle$ \cite{greven1994large}. Its sign determines the limiting behavior: for $\langle\log(k_i/r_i)\rangle>0$, the dynamics approach the deterministic limit, whereas for $\langle\log(k_i/r_i)\rangle < 0$, they converge to the exponential limit. This illustrates that even in the simplest setting, naive bias measures are insufficient, and that identifying the correct effective quantity requires a more refined description.}

Moreover, unbounded rates allow for transition rates that are macroscopically large, preventing FPT universality (see Supplement V).
\new{Instead, a more symmetric and robust classification in this case would also be the mean residual lifetime. Based on Eq.~(\ref{eq:MRL_limit}), it is transparent that the exponential limit arises if:
\begin{align}
    \lim_{N\to \infty}\frac{\lim_{t\to\infty}\langle\tau\rangle(t)}{\langle\tau\rangle}&= 1,
\end{align}
which reflects that the first-passage time must be effectively memoryless for all times in order to obtain the exponential limit.}

\subsection{First-passage time simulations for random networks}

\begin{figure}[t]
    \centering
    \includegraphics[width=0.95\linewidth]{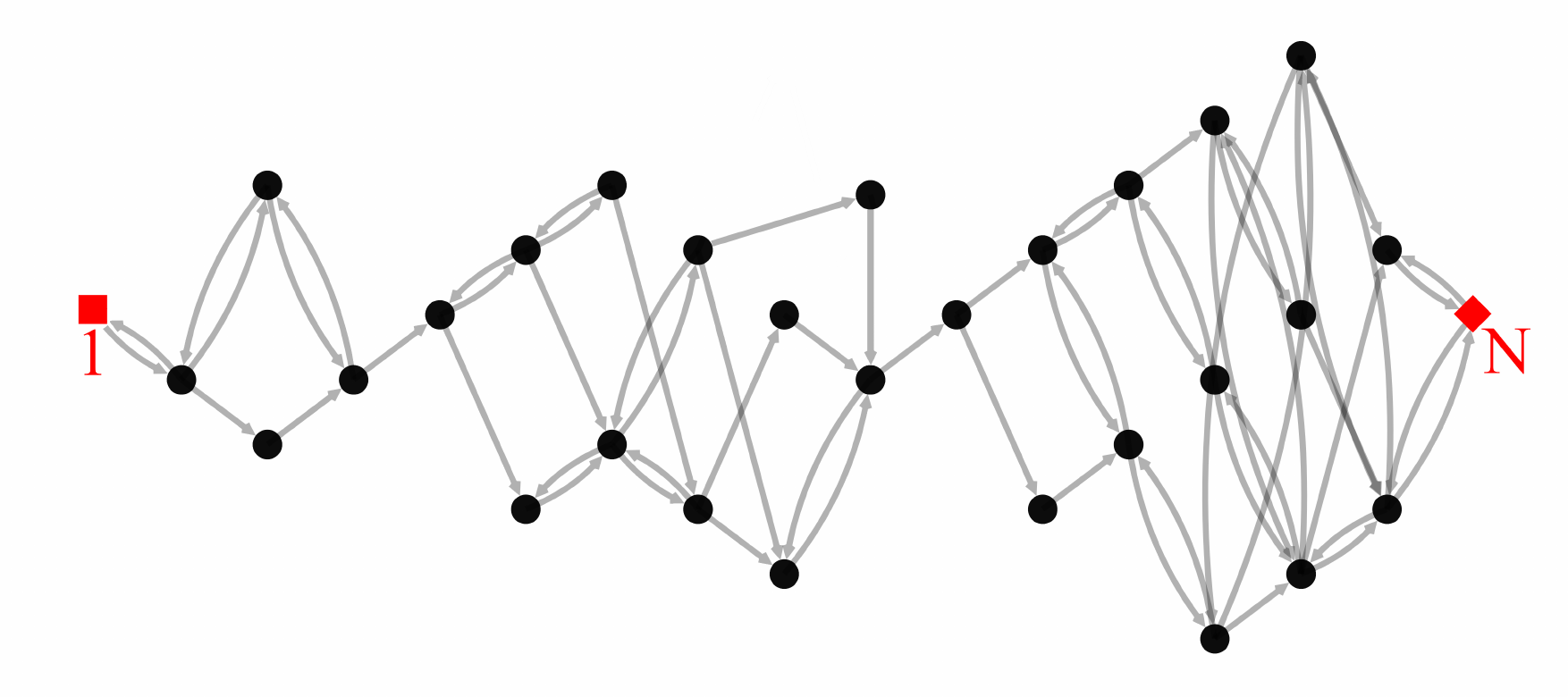}
    \caption{Example of a randomly generated graph with $N=30$ vertices based on the procedure introduced in \cite{voits2025generic} to obtain networks with clear starting and end points. Both the network
    topology and the rates are obtained by random processes.}
    \label{fig:network_example}
\end{figure}

To practically demonstrate how first-passage time distributions 
are related to the eigenvalue structure of the generator
matrix, we now consider random networks of increasing size $N$. 
We will consider both forward and backward biases as well as both reversible and irreversible dynamics. This requires networks with a source at vertex $1$ and a sink at vertex $N$ that are relatively far apart. Since random graphs generated by Erdős–Rényi-like models tend to give small graph distances in connected graphs, we employ the graph-generation procedure introduced in \cite{voits2025generic}, which yields random graphs such as the one shown in Fig. \ref{fig:network_example}. 

In the reversible case, all transitions are bidirectional and rates satisfy detailed balance. In the non-reversible case, directed edges are assigned independently, and backward transitions are included with a fixed probability $p$, so that detailed balance is generally broken. The rates for the edges $(i,j)$ are then assigned as:
\begin{align}
    k_{ij}=k_{ij}^0e^{b\frac{\text{d}(i,N)-\text{d}(j,N)}{2}},
\end{align}
where $d(i,N)$ and $d(j,N)$ denotes the graph distance from vertex 
$j$ and $i$ to the absorbing target $N$, respectively, and $b$ controls the bias toward ($b>0$) or away from ($b<0$) the target. The prefactors $k_{ij}^0$ are drawn from a uniform distribution,
\begin{align}
    k_{ij}^0\sim U([1-\alpha,1+\alpha]),
\end{align}
where $\alpha<1$ sets the variability of the rates.
In the reversible case, the prefactors are chosen symmetrically, $k_{ij}^0=k_{ji}^0$, while in the general case, they are drawn independently.

The stochastic dynamics are simulated using the Gillespie algorithm \cite{gillespie1977exact} and the eigenvalues of the corresponding Laplace matrices are computed numerically. Although the network structure and the rates are randomly generated, the resulting matrix retains the Laplacian structure, so its entries are not independent. Hence, simple limit laws for the eigenvalues of random matrices do not apply. More details on the simulations are given in Supplement VI. The code and the data are publicly available \cite{Github2026}.

\begin{figure*}[p]
    \centering
    \includegraphics[width=0.75\linewidth]{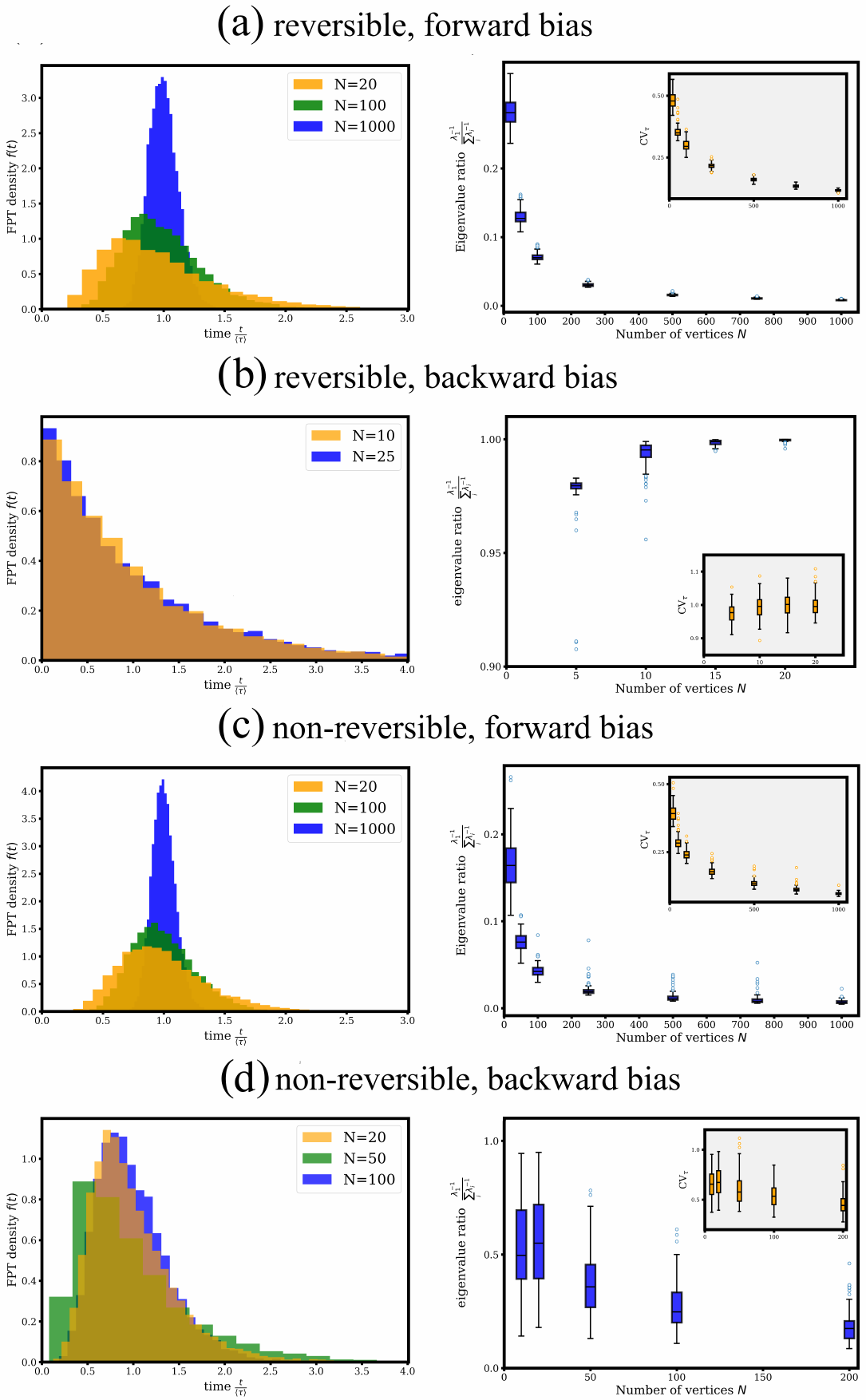}
    \caption{First-passage times and eigenvalue ratios for random networks of increasing size. (a) For reversible networks with forward bias, the FPT statistics approaches the deterministic limit as system size $N$ increases. This is reflected by a vanishing eigenvalue ratio (box plot)
    and a vanishing coefficient of variation (inset). (b) For reversible networks with backward bias, the FPT statistics approaches the exponential limit as system size $N$ increases, reflected in a dominant principal eigenvalue. The 
    coefficient of variation approaches $1$. (c) For non-reversible networks with forward bias, the FPT statistics approaches the deterministic limit as system size $N$ increases, similar to
    case (a). (d) For non-reversible networks with backward bias, the FPT limit is not generic. The eigenvalue ratio neither approaches $0$ nor $1$. The histograms show the FPT distribution for $10^5$ runs on a single network and the boxplots are obtained from $100$ networks on which $10^4$ first-passage times were generated.}
    \label{fig:fpts_random_nws}
\end{figure*}

Our results are shown in Fig.~\ref{fig:fpts_random_nws}.
The histograms display first-passage time distributions obtained from $10^5$ trajectories on representative networks with increasing vertex number $N$. The box plots show the dominance ratio of the principal eigenvalue,
\begin{align}
    R = \frac{\lambda_1^{-1}}{\sum_i \lambda_i^{-1}},
\end{align}
computed for $100$ independent network realizations for each $N$. The inset diagrams show the corresponding coefficients of variation,
\begin{align}
    \mathrm{CV}_\tau = \frac{\sigma_\tau}{\langle \tau \rangle},
\end{align}
obtained from $10^4$ first-passage time samples for each network.

The results for reversible networks of increasing size with a forward bias in the rates ($b=1$) are shown in Fig.~\ref{fig:fpts_random_nws}(a): as expected, the FPT distributions converge to the deterministic limit (histograms). The eigenvalue ratio (box plot) and the coefficient of variation of the first-passage time (inset) both vanish with increasing $N$. Fig.~\ref{fig:fpts_random_nws}(b) shows the limiting behavior for reversible networks with a backward bias $b=-1$: the exponential limit arises already for moderate system sizes (histograms). Accordingly, the principal eigenvalue dominates the spectrum reflected by the eigenvalue ratio (box plots) and coefficient of variation (inset) approaching unity. 

Fig.~\ref{fig:fpts_random_nws}(c) shows that 
for non-reversible networks, in which reverse transitions are present with probability $p=0.5$, the results for a forward bias in the rates ($b=1$) are qualitatively identical to the reversible case. Again, the deterministic limit arises (histograms) with vanishing eigenvalue ratio (histogram) and coefficient of variation (inset).
However, as shown in Fig.~\ref{fig:fpts_random_nws}(d), for non-reversible networks with a backward bias ($b=-1$), the limiting first-passage time distributions do not generically approach the exponential limit (histograms) and the eigenvalue ratio and the coefficient of variation do not converge to $1$. The reason for this is that unidirectional transitions at bottleneck positions in the network can lead to metastable clusters and checkpoint-like transitions between them. Hence, this case has to be handled with care and for networks with non-reversible transitions at crucial points, one should expect a non-generic limit.

\section{Discussion}
\label{sec:Discussion}

It has been reported before for kinetic proofreading networks that 
two simple limiting distributions arise in the limit of large networks
\cite{SimplicityComplexNW,munsky2009specificity}. This 
observation is surprising in light of the immense structural 
complexity of biochemical reaction networks \cite{tyson2020dynamical}.
It calls for a more thorough mathematical explanation, since FPTs in Markovian networks correspond to so-called phase-type distributions, which are dense in the class of distributions with non-negative support \cite{MatrixExpDistributions}. Consequently, they can, \textit{a priori}, approximate any possible shape of an FPT distribution arbitrarily well.

Here we have established that these limits
originate from the structure of the eigenspectrum of the generator matrix. We introduced a condition on the ratio of the weights of two-tree spanning forests, the macroscopic forest condition (Sec. \ref{subsec:Macroscopic_forest_condition}), which allowed us to relate the FPT cumulants to sums over the eigenvalues in the limit $N \to \infty$. From this relation, we deduced that if infinitely many eigenvalues contribute non-vanishingly to the sum, the FPT distribution converges to the deterministic limit (Sec. \ref{subsec:Deterministic_limit}). 

For strongly connected networks with locally finite outgoing rates, a finite conductance is sufficient to ensure convergence to this deterministic regime. For networks that are not strongly connected, the analysis must be performed separately for each strongly connected component. If the FPTs across components vanish compared to the global FPT as $N \to \infty$, the total FPT distribution becomes a sum of delta peaks corresponding to the distinct traversal paths through these components.
As a biological example, embryonic development may be viewed as a system undergoing many irreversible transitions. Although the mean completion time is sensitive to environmental factors such as temperature, the distribution of completion times remains narrowly concentrated around it \cite{chong2018temporal,rombouts2025mechanistic,jacobs2025beyond,jacobs2026understanding,jacobs2026understanding2}.

The exponential limit, on the other hand, arises when the generator matrix has a single dominant eigenvalue (Sec. \ref{subsec:Exponential_limit}). We demonstrated that, if the macroscopic network condition holds, showing that the coefficient of variation approaches unity is sufficient to guarantee convergence to an exponential distribution. Moreover, we established that, at least for reversible networks (i.e., those obeying detailed balance in the steady state), \new{this case permits a simple definition of a global backward bias in terms of the ratio of the MFPTs in forward and backward direction, yielding a relatively easy to check mathematical condition.} 
\new{However, as demonstrated in Sec.~\ref{subsec:Forward_not_sufficient}, a naive inversion of this condition is not enough to obtain the deterministic limit. For the rather simple case of a one-step master equation, a mathematically precise treatment requires considering the distribution of the logarithm of the ratios of forward and backward rates \cite{greven1994large}, which cannot serve as a general criterion for more complex networks, especially in the presence of irreversible transitions. 

A pathwise alternative would be to compare the expected number of jumps before absorption with the minimal graph distance to the target. For continuous-time Markov networks with weighted edges, however, such a measure should also account for the transition rates. A rate-sensitive characterization that captures both the deterministic and the exponential case for general networks can be obtained by considering the mean residual lifetime of the process, whose long-time limit is governed by the principal eigenvalue. If the mean residual lifetime becomes small at long times compared to the mean first-passage time, this implies that the contribution of the dominant eigenvalue vanishes, providing a formal condition for the deterministic limit. Conversely, if if the ratio of the long-time mean residual lifetime to the MFPT converges to $1$ for large networks, this results in the exponential limit.
This constitutes an interesting classification of the limiting behavior, not only because this one criterion covers both limits, but also because it yields a necessary signature for the two limits: in the deterministic case, the target is reached at a specific finite time, so $\langle\tau\rangle(t)\overset{t\to\infty}{\to} 0$, while the exponential distribution is memoryless, meaning that $\langle\tau\rangle(t)=\langle\tau\rangle=const.$ for all times. However, the mean residual lifetime seems less accessible than, for example, the mean first-passage times, and future research may focus on relating it to more structural properties of the network, such as the statistics of the transition rates.}

Simulations of first-passage times in random networks with random rates confirm that the deterministic (delta-like) limit emerges robustly in networks with a forward bias. In these cases, the ratio of the principal eigenvalue contribution to the total spectral weight vanishes with increasing system size, in agreement with the spectral analysis. Conversely, in reversible networks, a backward bias generically leads to dominance of a single eigenvalue and hence to exponential first-passage statistics, again as predicted by the spectral analysis.

In irreversible networks, however, the global structure of the graph can qualitatively alter this picture. A backward bias may generate metastable clusters connected by effectively irreversible transitions. These irreversible steps act as stochastic checkpoints, segmenting the dynamics into successive stages and thereby preventing generic spectral concentration. As a consequence, deviations from the simple exponential limit can arise despite the presence of a backward bias. A systematic classification of the backward-biased irreversible regime, in particular the role of metastable clustering and effective dynamical checkpoints, represents an interesting direction for future work.

Overall, we conclude that the universality of these limiting first-passage time distributions holds under clearly identifiable structural conditions. First, the large-network limit can yield universal behavior only if the process starts sufficiently far away from the target state. Otherwise, local non-macroscopic structures dominate. Thus, a condition such as the macroscopic forest condition is needed for universality. Second, irreversible transitions can break this universality: while a network composed of strongly connected microscopic subnetworks yields several delta peaks, if multiple \textit{macroscopic} strongly connected components exist with irreversible connections between them, the resulting FPT distribution becomes non-generic. One of the simplest examples would be a system with two macroscopic components, each individually leading to exponential FPTs, but connected by an irreversible transition. The total FPT distribution in the limit $N \to \infty$ then becomes a convolution of two exponentials — neither purely exponential nor deterministic. Finally, universality cannot be expected if arbitrarily large outgoing rates are allowed (compare the example in Supplement V).

Understanding how simple macroscopic laws emerge from the intricate dynamics of microscopic constituents lies at the heart of statistical physics. While this question was originally treated within the context of equilibrium systems, recent decades have seen remarkable progress in extending these principles to non-equilibrium settings \cite{seifert2012stochastic,jarzynski2017stochastic,tang2021topology,seifert2025stochastic}, such as living systems  \cite{england2013statistical, baouche2025first}. Earlier work has made 
large progress to identify universal limiting distributions for space-continuous processes,
often using Laplace-transforms \cite{benichou2010geometry,benichou2014first,godec2016universal,baravi2025solutions,baravi2025first}. 
Our work takes a complementary point of view by considering space-discrete Markovian networks and 
without assuming an underlying spatial or diffusive structure. 
In fact our results are valid also for different rates for
each transition. We show that the emergence of simple FPT-distributions can be understood in terms of the eigenvalue distribution of the generator, providing a unifying classification that extends the concept of universality in first-passage processes from geometry- and transport-controlled kinetics to general networks with a large parameter space such as biochemical reaction networks. In this framework, exponential behavior arises from the dominance of a single eigenvalue, consistent with previous observations in diffusion-controlled systems. In contrast, when many eigenvalues contribute comparably, the FPT-distribution approaches a quasi-deterministic distribution around its mean, which we characterize here as a distinct generic limit.

More broadly, our results highlight that even in high-dimensional Markovian networks, the interplay between spectral properties of the generator and the underlying graph structure governs the emergence of simple limiting first-passage times. For large and complex networks, approaches based on spectral theory \cite{hartich2019extreme,hartich2019interlacing} and complementary graph-theoretical formalisms \cite{nam2025algebraic, voits2025generic} therefore provide a powerful unified framework for uncovering general mechanisms that remain hidden in the enormous microscopic parameter space. These insights set the stage for extending the present results to broader classes of stochastic systems and for exploring universal behavior beyond the regimes addressed here.

\begin{acknowledgments}

JBV thanks the German Academic Scholarship Foundation (Studienstiftung des Deutschen Volkes) for support. 
We thank David Geldbach and Eli Barkai for helpful comments.
We also acknowledge support by the Max Planck School Matter to Life funded by the Dieter Schwarz Foundation and the Max Planck Society. 

\end{acknowledgments}



\vspace{2cm}


\providecommand{\noopsort}[1]{}\providecommand{\singleletter}[1]{#1}%

\end{document}


\title{Supplemental Material for\\
Emergence of generic first-passage time distributions for large Markovian networks}

\author{Julian B. Voits\textsuperscript{1}}
\author{Ulrich S. Schwarz\textsuperscript{1,2}}%
 \email{Corresponding author: schwarz@thphys.uni-heidelberg.de}
\affiliation{%
\textsuperscript{1}Institute for Theoretical Physics, University of Heidelberg, Germany\\ \textsuperscript{2}BioQuant-Center for Quantitative Biology, University of Heidelberg, Germany
}%
\date{\today}

\maketitle

\tableofcontents
\bigskip

This Supplemental Material provides the mathematical details on the results presented in the main text of
\textit{“Emergence of generic first-passage time distributions
for large Markovian networks”}. All notation follows that of the main paper.

\section{All-Minors Matrix-Tree Theorem}
\noindent The All-Minors Matrix-Tree Theorem states the following:

\textit{Let $\hat{K}$ be the (in-degree) Laplacian and let $I,J\subseteq\{1,...,N\}$ with $|I|=|J|=n$. Denote by $\hat{K}_{I,J}$ the matrix obtained by deleting the rows indexed by $I$ and the columns indexed by $J$. Then:
\begin{align}
        \det\hat{K}_{I,J}=(-1)^{\sum_{i\in I}i+\sum_{j\in J}j}\sum_{\mathcal{F}^{I\to J}}w(\mathcal{F}),
\end{align}
where each tree in $\mathcal{F}^{I\to J}$ contains exactly one vertex from $I$ and one from $J$ such that the vertex in $I$ and the vertex in $J$ on the same tree are connected in one direction.}

Note that, if $N$ is an absorbing vertex, one has as an immediate corollary:
\begin{align}
        \det\hat{K}_{\{m\},\{N\}}&=(-1)^{m+N}\sum_{\mathcal{F}^{m\to N}}w(\mathcal{F})=(-1)^{m+N}\sum_{\mathcal{T}_{[N]}}w(\mathcal{T})=\det{K},
\end{align}
so the determinant of any minor of $\hat{K}$ obtained from deleting the $N$-th column and any row must be the same up to a sign.  

The theorem was first proven in full generality by Chaiken \cite{chaiken1982combinatorial}. Its application to the mean first-passage time goes back to Chebotarev\cite{chebotarev2007graph,pitman2018tree}, while the application to higher moments is a relatively recent extension by Nam and Gunawardena\cite{NamPhD,nam2023linear,nam2025algebraic}.

\section{Macroscopic forest condition}
In the main text, we stated the macroscopic forest condition, which was based on the observation that the mean first-passage time $\langle\tau_m\rangle$ starting at vertex $m$ is related to the sum of eigenvalues of of the generator matrix as follows:

\begin{align}
        \frac{1}{1+\frac{\sum_{j=1}^{N-1}\sum_{\mathcal{F}^{m\to j}_{[j,N]}}w(\mathcal{F})}{\sum_{j=1}^{N-1}\sum_{\mathcal{F}^{m\to N}_{[j,N]}}w(\mathcal{F})}}&=\frac{\sum_{j=1}^{N-1}\sum_{\mathcal{F}^{m\to N}_{[j,N]}}w(\mathcal{F})}{\sum_{j=1}^{N-1}\sum_{\mathcal{F}_{[j,N]}}w(\mathcal{F})}\\ &=1-\frac{\sum_{j=1}^{N-1}\sum_{\mathcal{F}^{m\to j}_{[j,N]}}w(\mathcal{F})}{\sum_{j=1}^{N-1}\sum_{\mathcal{F}_{[j,N]}}w(\mathcal{F})}\\ &=1-\frac{\langle\tau_m\rangle}{\text{tr}(M)}\\&=1-\frac{\langle\tau_m\rangle}{\sum_{i=1}^{N-1}\frac{1}{\lambda_i}},
\end{align}
using that $\text{tr}(M)=\frac{\sum_{j=1}^{N-1}\sum_{\mathcal{F}_{[j,N]}}w(\mathcal{F})}{\sum_{\mathcal{T}_{[N]}}w(\mathcal{F})}$. 

Using that the quotient of two sums with positive summands is bounded from above by the maximum quotient of a summand in the numerator and in the denominator, one has:
\begin{align}
    r:=&\frac{\sum_{j=1}^{N-1}\sum_{\mathcal{F}^{m\to N}_{[j,N]}}w(\mathcal{F})}{\sum_{j=1}^{N-1}\sum_{\mathcal{F}^{m\to j}_{[j,N]}}w(\mathcal{F})}\\\leq&\max_{j=1,...,N-1}\frac{\sum_{\mathcal{F}^{m\to N}_{[j,N]}}w(\mathcal{F})}{\sum_{\mathcal{F}^{m\to j}_{[j,N]}}w(\mathcal{F})}\\=&\max_{j=1,...,N-1}\frac{\sum_{\mathcal{F}^{m\to N}_{[j,N]}}w(\mathcal{F})}{\sum_{\mathcal{F}_{[j,N]}}w(\mathcal{F})-\sum_{\mathcal{F}^{m\to N}_{[j,N]}}w(\mathcal{F})}\\ =&\max_{j=1,...,N-1}\frac{\frac{\sum_{\mathcal{F}^{m\to N}_{[j,N]}}w(\mathcal{F})}{\sum_{\mathcal{F}_{[j,N]}}w(\mathcal{F})}}{1-\frac{\sum_{\mathcal{F}^{m\to N}_{[j,N]}}w(\mathcal{F})}{\sum_{\mathcal{F}_{[j,N]}}w(\mathcal{F})}}\\ =&\max_{j=1,\ldots,N-1} \frac{p_{m\to\{j,N\}}(N)}{1-p_{m\to\{j,N\}}(N)},
\end{align}
where the last line follows from the indentity for the splitting probability\cite{nam2023linear}:
\begin{align}
    p_{m\to\{j,N\}}(N)=\frac{\sum_{\mathcal{F}^{m\to N}_{[j,N]}}w(\mathcal{F})}{\sum_{\mathcal{F}_{[j,N]}}w(\mathcal{F})}
\end{align}

\begin{figure*}[tbh]
    \centering
    \includegraphics[width=1.0\textwidth]{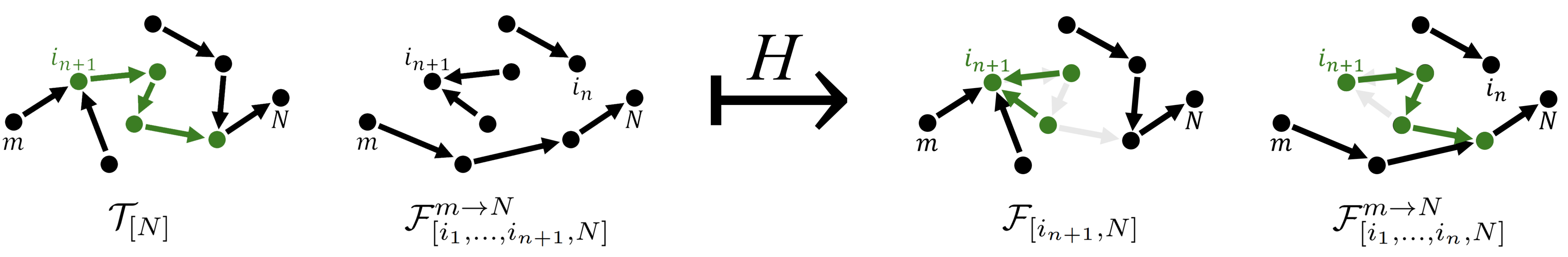}
    \caption{Illustration of the auxiliary map that maps a tuple of a spanning tree rooted at $N$ and an $n+2$ tree spanning forest rooted at $i_1,..,i_{n+1}, N$ and containing an $m\to N$ path to a tuple of a two tree spanning forests with roots $i_{n+1}$ and $N$ and an $n+1$ tree spanning forest rooted at $i_1,..,i_{n}, N$ and containing an $m\to N$ path.}
    \label{fig:H_map}
\end{figure*}

Now consider the logarithm of the Laplace transform $\mathcal{L}{f}(s)$ of the FPT density and make the replacement $s\to\frac{s}{\langle\tau_m\rangle}$:
\begin{align}
    \ln(\mathcal{L}{f}(s))=&-\sum_{i=1}^{N-1}\ln\big(1+\frac{s}{\lambda_i\langle\tau_m\rangle}\big)\\&+\ln\sum_{n=0}^{N-1}\frac{1}{\langle\tau_m\rangle^n}\sum_{i_1,...,i_n} \sum _{\mathcal{F}_{[i_1,...,i_n,N]}^{m\rightarrow N}}w(\mathcal{F})s^n\\ &=-\sum_{n=1}^\infty\frac{(-s)^n}{n}\frac{1}{\langle\tau_m\rangle^n}\sum_{i=1}^{N-1}\frac{1}{\lambda_i^n}\\ &+\ln\sum_{n=0}^{N-1}\frac{\sum_{i_1,...,i_n} \sum _{\mathcal{F}_{[i_1,...,i_n,N]}^{m\rightarrow N}}w(\mathcal{F})}{\langle\tau_m\rangle^n\sum_{\mathcal{T}_{[N]}}w(\mathcal{T})}s^n.
\end{align}
The first term is what we want, so we need to show that the second term vanishes as $N\to\infty$. To that end, define the sequence:
\begin{align}
    r_n:=&\frac{\sum_{i_1,...,i_n} \sum _{\mathcal{F}_{[i_1,...,i_n,N]}^{m\rightarrow N}}w(\mathcal{F})}{\langle\tau_m\rangle^n\sum_{\mathcal{T}_{[N]}}w(\mathcal{T})} =\frac{\sum_{i_1,...,i_n} \sum _{\mathcal{F}_{[i_1,...,i_n,N]}^{m\rightarrow N}}w(\mathcal{F})\big(\sum_{\mathcal{T}_{[N]}}w(\mathcal{T})\big)^{n-1}}{\big(\sum_{j=1}^{N-1}\sum_{\mathcal{F}^{m\to j}_{[j,N]}}w(\mathcal{F})\big)^n}.
\end{align}
Next, we show that:
\begin{align}
    \frac{r_{n+1}}{r_n(1+r)}=\frac{\sum_{i_1...i_{n+1}} \sum _{\mathcal{F}_{[i_1...i_{n+1},N]}^{m\rightarrow N}}w(\mathcal{F})\sum_{\mathcal{T}_{[N]}}w(\mathcal{T})}{\sum_{i_1...i_n} \sum _{\mathcal{F}_{[i_1...i_n,N]}^{m\rightarrow N}}w(\mathcal{F})\sum_{j=1}^{N-1}\sum_{\mathcal{F}_{[j,N]}}w(\mathcal{F})}\leq \frac{1}{n+1}.
\end{align}
If we show that the denominator contains all terms in the numerator $n+1$-times, the claim follows immediately. To that end, consider the auxiliary map
\begin{align*}
    H:T_{[N]}\times F^{m\to N}_{[i_1,...,i_{n+1},N]}&\to F_{[i_{n+1},N]}\times F^{m\to N}_{[i_1,...,i_{n},N]} \\ (\mathcal{T}_{[N]},\mathcal{F}^{m\to N}_{[i_1,...,i_{n+1},N]})&\mapsto (\mathcal{F}_{[i_{n+1},N]},\mathcal{F}^{m\to N}_{[i_1,...,i_n,N]}),\label{eq:auxiliary_map}
\end{align*}
where $T_{[N]}$ are the trees rooted at $N$, and $F$ are the forests with the corresponding roots, defined by choosing the part of the unique $i_{n+1}\to N$ path in $\mathcal{T}_{[N]}$ until it first reaches a vertex of either the $i_1,...i_N$-trees or the $N$-tree in $\mathcal{F}^{m\to N}_{[i_1,...,i_{n+1},N]}$ (if $i_{n+1}$ is on the $m\to N$ path, take the whole $i_{n+1}\to N$ ). Take all the edges in $\mathcal{F}^{m\to N}_{[i_1,...,i_{n+1},N]}$ along the path  and exchange the two subgraphs. This is illustrated in Fig. \ref{fig:H_map}.

This is in fact a well defined function since it exchanges edges from $\mathcal{T}_{[N]}$ with edges from the $i_{n+1}$-tree of $\mathcal{F}^{m\to N}_{[i_1,...,i_{n+1},N]}$, which reduces the edge count of the first graph by one and increases it for the second graph by one. In the first entry, $i_{n+1}$ does no longer have an outgoing edge because it did not have one in $\mathcal{F}^{m\to N}_{[i_1,...,i_{n+1},N]}$. Except for $N$, all other vertices have an outgoing edge and since the edge count is $N-2$, they must have exactly one outgoing edge. Hence, this is indeed a two tree spanning forest rooted at $N$. 
The second entry overall gained an edge and only the outgoing edges of the vertices on the $i_{n+1}$-tree could have been changed, so  and also the $m\to N$ path is unchanged and $i_1,...,i_n,N$ are still roots. All other vertices now have an outgoing edge and since there are now $N-(n+2)$ edges, they must have exactly one showing that this is an $n+1$ tree spanning forest rooted at $N$.

Observe that by construction of the map $\mathcal{F}^{m\to N}_{[i_1,...,i_{n},N]}$ does not contain an $m\to i_{n+1}\to N$. Hence, $H$ is not surjective. But $H$ is in fact injective, which can be seen by restricting its range to its image and noting that then the map can be inverted by following the $i_{n+1}\to i_n$ or $i_{n+1}\to N$ path in $\mathcal{F}_{[i_1,...,i_n,N]}$ until it reaches the $N$-tree in $\mathcal{F}^{m\to N}_{[i_{n+1},N]}$ (which it eventually does since there must be an $i_n\to N$ path if there is no $i_{n+1}\to N$ path) and by exchanging the corresponding edges again with $\mathcal{F}_{[i_{n+1},N]}$, recovering the original tree-forest tuple. 

The choice of $i_{n+1}$ was arbitrary, one could have chosen any of $i_1,...,i_{n+1}$, meaning that the denominator contains in fact $n+1$ corresponding tuples. Thus, we conclude that $\frac{r_{n+1}}{r_n}\leq \frac{1+r_1}{n+1}$. So, $r_n\leq \frac{1}{n+1}(1+r_1)^{n-1}r_1$ for all $n\neq 0$ and in particular:
\begin{align}
    1\leq\sum_{n=0}^{N-1}r_ns^n \leq  1+\frac{r_1}{1+r_1}\sum_{n=1}^\infty\frac{ (1+r_1)^{n}}{n!}s^n=1+\frac{r_1}{1+r_1}(e^{(1+r_1)s}-1),
\end{align} 
since $r_0=1$. If $r_1=0$, one can conclude that $\sum_{n=0}^{N-1}r_ns^n$ converges to $1$. Hence, one has:
\begin{align}
    \lim_{N\to\infty}\ln(\mathcal{L}{f}(s))=-\sum_{n=1}^\infty\frac{(-s)^n}{n}\lim_{N\to\infty}\frac{\sum_{i=1}^{N-1}\frac{1}{\lambda_i^n}}{\langle\tau_m\rangle^n}.
\end{align}

So, if the previous condition holds, one gets for the cumulants:
\begin{align}
    \lim_{N\to\infty}\kappa_n \Big(\frac{t}{\langle\tau_m\rangle}\Big) =& (n-1)!\lim_{N\to\infty}\frac{\sum_{i=1}^{N-1}\frac{1}{\lambda_i^n}}{\langle\tau_m\rangle^n}=(n-1)!\lim_{N\to\infty}\frac{\sum_{i=1}^{N-1}\frac{1}{\lambda_i^n}}{\Big(\sum_{i=1}^{N-1}\frac{1}{\lambda_i}\Big)^n},
\end{align}
since the mean is the first cumulant.

\section{Deterministic limit}
As stated in the main text, it is sufficient to show that $\text{Re}(\lambda_i)\geq M>0$ and $\sum_{i=1}^N\frac{1}{\lambda_i}\overset{N\to\infty}{\longrightarrow}\infty$ to conclude that:
\begin{align}
    \frac{\sum_{i=1}^N\frac{1}{\lambda_i^2}}{(\sum_{i=1}^N\frac{1}{\lambda_i})^2}\overset{N\to\infty}{\to}0.
\end{align}

The reason for this is that the characteristic polynomial:
\begin{align}
    \prod_{i=1}^{N-1}(s+\lambda_i)=\sum_{\mathcal{T}_{s,[N]}}w(\mathcal{T})
\end{align}
has only real coefficients. Hence, if the eigenvalues are complex, they must occur in conjugate pairs and one has:
\begin{align}
    \sum_{i=1}^N\frac{1}{\lambda_i}=\sum_{i=1}^N\frac{\text{Re}\lambda_i}{|\lambda_i|^2}.
\end{align}
With that, one finds:
\begin{align}
    \Big|\sum_{i=1}^N\frac{1}{\lambda_i^2}\Big|\leq\sum_{i=1}^N\frac{1}{|\lambda_i|^2}\leq\frac{1}{M}\sum_{i=1}^N\frac{\text{Re}(\lambda_i)}{|\lambda_i|^2} =\frac{1}{M}\sum_{i=1}^N\frac{1}{\lambda_i},
\end{align}
implying that:
\begin{align}
    \frac{\Big|\sum_{i=1}^N\frac{1}{\lambda_i^2}\Big|}{\Big(\sum_{i=1}^N\frac{1}{\lambda_i}\Big)^2}\leq\underbrace{\frac{1}{M\sum_{i=1}^N\frac{1}{\lambda_i}}}_{\to 0}.
\end{align}

\section{Exponential limit}
For the exponential limit, we argued in the main text that it is sufficient to show that the coefficient of variation converges to $1$, $\lim_{N\to\infty}\frac{\sigma_m}{\langle\tau_m\rangle}=1$, if the macroscopic forest condition holds with $r=0$ since then also holds $\lim_{N\to\infty}\kappa_{n}(\frac{t}{\langle\tau_m\rangle})=(n-1)!$ $\forall n$.
To see this, define $w_j:=\frac{\frac{1}{\lambda_j}}{\sum_{l=1}^{N-1}\frac{1}{\lambda_j}}=a_j+ib_j$, with $0\leq a_j\leq 1$, monotonically decreasing and the $w_j$s occurring in conjugated pairs if $b_j\neq0$, so $\sum_{j=1}^{N-1}w_j=\sum_{j=1}^{N-1}a_j=1$. This means in particular that $\sum_{j=1}^{N-1}a_j^2\leq1$ therefore it holds:
    \begin{align}
        &1=\lim_{N\to\infty}\frac{\sigma_m^2}{\langle\tau_m\rangle^2}=\lim_{N\to\infty}\sum_{j=1}^{N-1}w_j^2=\lim_{N\to\infty}\sum_{j=1}^{N-1}(a_j^2-b_j^2)\\ \Rightarrow & 0=1-\lim_{N\to\infty}\sum_{j=1}^{N-1}w_j^2=\lim_{N\to\infty}(\underbrace{1-\sum_{j=1}^{N-1}a_j^2}_{\geq0})+\lim_{N\to\infty}\underbrace{\sum_{j=1}^{N-1}b_j^2}_{\geq0}.
    \end{align}
    which requires $\lim_{N\to\infty}\sum_{j=1}^{N-1}a_j^2=1$ and $\lim_{N\to\infty}\sum_{j=1}^{N-1}b_j^2=0$, so $b_j\overset{N\to\infty}{\to}0$ $\forall j$. Since:
    \begin{align}
        \sum_{j=1}^{N-1}a_j^2\leq a_1\sum_{j=1}^{N-1}a_j=a_1\leq 1,
    \end{align}
    so $w_1=a_1\overset{N\to\infty}{\to} 1$ and $\lim_{N\to\infty}\sum_{j=2}^{N-1}a_j=0$. This also means that $\lim_{N\to\infty}\sum_{j=2}^{N-1}a_j^2=0$ and in particular, $a_j\to 0$ for $j\geq 2$. Hence, $\max_{2\leq j\leq N-1}|w_j|=\max_{2\leq j\leq N-1}\sqrt{a_j^2+b_j^2}\overset{N\to\infty}{\to}0 $, and for $n\geq 3$ holds:
    \begin{align}
        0&\leq \Big|\sum_{j=2}^{N-1}w_j^n\Big|\leq \sum_{j=2}^{N-1}|w_j|^n\leq \big(\max_{2\leq j\leq N-1}|w_j|\big)^{n-2} \sum_{j=2}^{N-1}|w_j|^2=\underbrace{\big(\max_{2\leq j\leq N-1}|w_j|\big)^{n-2}}_{\to 0}\underbrace{ \sum_{j=2}^{N-1}(a_j^2+b_j^2)}_{\to 0}.
    \end{align}
    So $0=\lim_{N\to\infty}\sum_{j=2}^{N-1}w_j^n$ and one can conclude that:
    \begin{align}
        \lim_{N\to\infty}\frac{\kappa_{n,m}}{\langle\tau_m\rangle^n}&=(n-1)!\lim_{N\to\infty}\sum_{j=1}^{N-1}w_j^n=(n-1)!\Big(1+\lim_{N\to\infty}\sum_{j=2}^{N-1}w_j^n\Big)=(n-1)!.
    \end{align}

Moreover, we gave a sufficient condition based on the backward bias interpretation for the exponential limit under the assumption that the network is reversible, meaning that the equilibrium distribution of $\hat{K}$ satisfies detailed balance: let $\langle\tau\rangle_{i\to j}$ denote the MFPT from $i$ to $j$ and let $\langle\tau\rangle_{m\to N}=\max_{i}\langle\tau\rangle_{i\to N}$. If:
\begin{align}
    \frac{ \langle\tau\rangle_{N\to m}}{\langle\tau\rangle_{m\to N}}\overset{N\to\infty}{\to}0,
\end{align}
and $ \lim_{N\to\infty}\frac{\sum_{j=1}^{N-1}\sum_{\mathcal{F}^{N\to m}_{[j,m]}}w(\mathcal{F})}{\sum_{j=1, j\neq m}^{N}\sum_{\mathcal{F}^{N\to j}_{[j,m]}}w(\mathcal{F})}=r< \infty$, then $\lim_{N\to\infty}\kappa_{n}(\frac{t}{\langle\tau_m\rangle})=(n-1)!$, so the first-passage time distribution for $m\to N$ converges to an exponential distribution.

The reversibility condition means that $\hat{K}$ is similar to a symmetric matrix $P\hat{K}P^{-1}$, where $P=\text{diag}(\sqrt{p_{s,1}},...,\sqrt{p_{s,N}})$, and its eigenvalues are real, as argued in the main text.
Observe that then also $L_\alpha :=\hat{K}+\text{diag}(0,...,0,\alpha)$ is similar to a symmetric matrix $PL_\alpha P^{-1}$ $\forall \alpha$ and hence, only has real roots. We can use the arguments presented in\cite{fisk2005very} as follows:
Consider the characteristic polynomial of $L_\alpha$:
\begin{align}
    \det(L_\alpha-\lambda\mathds{1})&=\det\begin{pmatrix}
    K -\lambda\mathds{1}& \vline & u \\
    \hline
    v^T & \vline & \sum_jk_{Nj}+\alpha-\lambda
    \end{pmatrix}\\ &=\det\begin{pmatrix}
    K -\lambda\mathds{1}& \vline & u \\
    \hline
    v^T & \vline & \sum_jk_{Nj}-\lambda
    \end{pmatrix} \\&+\det\begin{pmatrix}
    K -\lambda\mathds{1}& \vline & u \\
    \hline
    0 & \vline &\alpha
    \end{pmatrix}\\ &=\underbrace{\det(\hat{K}-\lambda\mathds{1})}_{=:f(\lambda)}+\alpha\underbrace{\det(K -\lambda\mathds{1})}_{:=g(\lambda)},
\end{align}
and we know that the left-hand side only has real roots from the previous consideration. 
There is a theorem stating that the roots of two polynomials $f$ and $g$ interlace if and only if $f+\alpha g$ has only real roots for all $\alpha\in\mathbb{R}$, which is precisely the case here.

The choice of the terminal vertex $N$ was arbitrary. One can analogously obtain the first-passage time density to $m$ by considering the $m$-th principal minor, and in particular the mean $\langle\tau\rangle_{i\to m}$ for any initial state $i$. Denote the eigenvalues of the $m$-th principal minor of $\hat{K}$ as $|\lambda_1^{(m)}|\leq|\lambda_2^{(m)}|\leq...\leq |\lambda_{N-1}^{(m)}|$ and of the $N$-th principal minor as $|\lambda_1^{(N)}|\leq|\lambda_2^{(N)}|\leq...\leq |\lambda_{N-1}^{(N)}|$ and the eigenvalues of $\hat{K}$ as $0=|\mu_1|\leq|\mu_2|\leq ...\leq |\mu_N|$.
As shown before, the eigenvalues are interlacing, meaning that:
\begin{align}
    0=\mu_1\leq\lambda_1^{(l)}\leq ...\leq\mu_{N-1}\leq \lambda_{N-1}^{(l)}\leq\mu_N,
\end{align}
for $l=1,...,N$. In particular, this means that:
\begin{align}
    \lambda_{j-1}^{(l_1)}&\leq\lambda_j^{(l_2)}\leq\lambda_{j+1}^{(l_1)}\Rightarrow \frac{1}{\lambda_{j-1}^{(l_1)}}\geq\frac{1}{\lambda_j^{(l_2)}}\geq\frac{1}{\lambda_{j+1}^{(l_1)}}.
\end{align}
With the assumption that $ \lim_{N\to\infty}\frac{\sum_{j=1}^{N}\sum_{\mathcal{F}^{i\to m}_{[j,m]}}w(\mathcal{F})}{\sum_{j=1, j\neq m}^{N}\sum_{\mathcal{F}^{i\to j}_{[j,m]}}w(\mathcal{F})}=r< \infty$, one has:
\begin{align}
    \lim_{N\to\infty}\langle\tau\rangle_{N\to m}&=\frac{1}{1+r}\lim_{N\to\infty}\sum_{i=1}^{N-1}\frac{1}{\lambda_i^{(m)}}\geq\frac{1}{1+r}\lim_{N\to\infty}\sum_{i=2}^{N-1}\frac{1}{\lambda_i^{(N)}}\\\Rightarrow\lim_{N\to\infty} \frac{\langle\tau\rangle_{N\to m}}{\langle\tau\rangle_{m\to N}}&\geq\frac{1}{1+r}\lim_{N\to\infty}\frac{\sum_{i=2}^{N-1}\frac{1}{\lambda_i^{(N)}}}{\sum_{i=1}^{N-1}\frac{1}{\lambda_i^{(N)}}},
\end{align}
using that $\langle\tau\rangle_{m\to N}\leq \sum_{i=1}^{N-1}\frac{1}{\lambda_i^{(N)}}$. Now, if $\lim_{\to\infty}\frac{\langle\tau\rangle_{N\to m}}{\langle\tau\rangle_{m\to N}}=0$ it follows immediately that:
\begin{align}
    \lim_{N\to\infty}\frac{\sum_{n=2}^N\frac{1}{\lambda_n^{(N)}}}{\frac{1}{\lambda_1^{(N)}}} = 0.
\end{align}
And since $\frac{1}{\lambda_1}\leq \langle\tau\rangle_{m\to N}\leq\sum_{i=1}^{N-1}\frac{1}{\lambda_i}$, this means:
\begin{align}
    \lim_{N\to\infty}\frac{\langle\tau\rangle_{m\to N}}{\sum_{i=1}^{N-1}\frac{1}{\lambda_i}}=\lim_{N\to\infty}\frac{\langle\tau\rangle_{m\to N}}{\frac{1}{\lambda_1}}=1.
\end{align}
As seen before, this implies that $\lim_{N\to\infty}\kappa_n \Big(\frac{t}{\langle\tau_m\rangle}\Big) =(n-1)!\lim_{N\to\infty}\frac{\sum_{i=1}^{N-1}\frac{1}{\lambda_i^n}}{\Big(\sum_{i=1}^{N-1}\frac{1}{\lambda_i}\Big)^n}$ and since:
\begin{align}
    0\leq \frac{\sum_{j=2}^{N-1}\frac{1}{\lambda_j^n}}{\frac{1}{\lambda_1^{n}}}\leq\underbrace{\frac{\lambda_1 ^{n-1}}{\lambda_2 ^{n-1}}}_{\leq 1} \underbrace{\frac{\sum_{j=2}^{N-1}\frac{1}{\lambda_j}}{\frac{1}{\lambda_1}}}_{\to 0}
\end{align}
Hence, $\lim_{N\to\infty}\frac{\sum_{j=1}^{N-1}\frac{1}{\lambda_j^n}}{\frac{1}{\lambda_1^{n}}}=0$ $\forall n$ so that indeed $\lim_{N\to\infty}\kappa_n \Big(\frac{t}{\langle\tau_m\rangle}\Big)=(n-1)!$.

\section{Explicit calculation of the first-passage time limit of a linear chain}
Consider a linear chain of irreversible transitions starting at $m=1$, i.e., a one-step master equation with $r_i=0$. Then the first passage time is simply the convolution of the exponentially distribted jumping times at each site:
\begin{align}
    f(t)=\prod_{i=1}^{N-1}k_i\Big(e^{-k_1t}\ast e^{-k_2t}\ast...\ast e^{-k_{N-1}t}\Big)(t),
\end{align}
which turns into a a product in Laplace space:
\begin{align}
    \mathcal{L}{f}(s)=\prod_{i=1}^N\frac{k_i}{s+k_i}=\prod_{i=1}^N\frac{1}{1+\frac{s}{k_i}}.
\end{align}
One then gets:
\begin{align}
    \langle \tau \rangle &=-\frac{\text{d}\mathcal{L}{f}}{\text{d}s}(s)\Big|_{s=0}=\sum_{i=1}^N\frac{1}{k_i}\\\langle \tau^2 \rangle &=\frac{\text{d}^2\mathcal{L}{f}}{\text{d}s^2}(s)\Big|_{s=0}=\Big(\sum_{i=1}^N\frac{1}{k_i}\Big)^2+\sum_{i=1}^N\frac{1}{k_i^2}
\end{align}
Normalizing time $\frac{t}{\langle\tau\rangle}\to t$ means that in the Laplace transform, the replacement $s\to\frac{s}{\langle\tau\rangle }$ has to be made:
\begin{align}
    \mathcal{L}{f}(s)=\prod_{i=1}^N\frac{1}{1+\frac{s}{\langle\tau\rangle k_i}}.
\end{align}
Now consider the first derivative of this expression:
\begin{align}
    \frac{\text{d}\mathcal{L}{f}}{\text{d}s}(s)&=-\sum_{j=1}^N\frac{1}{\langle\tau\rangle k_j}\frac{1}{1+\frac{s}{\langle\tau\rangle k_j}}\prod_{i=1}^N\frac{1}{1+\frac{s}{\langle\tau\rangle k_i}}\\ &=-\sum_{j=1}^N\frac{1}{\langle\tau\rangle k_j}\frac{1}{1+\frac{s}{\langle\tau\rangle k_j}}\mathcal{L}{f}(s).
\end{align}
If the $k_i$ satisfy  $\frac{\sum_{i=1}^N\frac{1}{k_i^2}}{\Big (\sum_{i=1}^N\frac{1}{k_i}\Big )^2}\to 0$, one gets:
\begin{align}
    1&=\frac{1}{\langle\tau\rangle }\sum_{j=1}^N\frac{1}{k_j}\\&\geq\frac{1}{\langle\tau\rangle }\sum_{j=1}^N\frac{1}{k_j}\frac{1}{1+\frac{s}{\langle\tau\rangle k_j}}\\&\geq \Big ( \sum_{j=1}^N\frac{1}{k_j\langle\tau\rangle}(1+\frac{s}{\langle\tau\rangle k_j})\Big )^{-1}\\&=\underbrace{\frac{1}{1+s\frac{\sum_{i=1}^N\frac{1}{k_i^2}}{\langle \tau\rangle ^2}}}_{\to 1}, 
\end{align}
applying Jensen's inequality, since $f(x)=x^{-1}$ is convex.

Hence, for $N\to\infty$, one has:
\begin{align}
    \frac{\text{d}\mathcal{L}{f}}{\text{d}s}(s)=-\mathcal{L}{f}(s)\implies f(s)=e^{-s}.
\end{align}
Note that $\mathcal{L}{f}(s=0)=1$, for normalization. In particular, this means that $f(t)=\delta(t-1)$, so in the limit of a large chain, the FPT distribution becomes deterministic on the scale of $\langle\tau\rangle$.

But observe that this need not hold true if the rates are unbounded. For the choice $k_i=i^2$, one gets for the coefficient of variation:
\begin{align}
    \lim_{N\to\infty}\text{CV}^2=\lim_{N\to\infty}\frac{\sum_{i=1}^N\frac{1}{k_i^2}}{\Big (\sum_{i=1}^N\frac{1}{k_i}\Big )^2}=\frac{4}{15},
\end{align}
meaning that one does not recover the deterministic limit but the limiting distribution keeps a finite width.

\section{Random networks}

Investigating the first-passage times for increasing system sizes required random networks on $N$ vertices with source $1$ and sink $N$ separated by a relatively large graph distance. Therefore, Erdos-Renyi models and similar random graphs were not approriate as they lead to relatively small distances in connected graphs. Instead, the networks were generated using the hierarchical construction introduced in Ref.~\cite{voits2025generic}: a directed layered skeleton graph on $\frac{N}{10}$ vertices is first constructed, ensuring a large graph distance between source and sink. Each vertex in the skeleton graph is then replaced by an independently generated layered subgraph, and edges in a block are attached according to the skeleton connectivity. Reverse transitions are added independently for all present edges with probability $p$. The reversible case requires $p=1$, while we chose $p=0.5$ to explore the effects of non-reversibility.
The transition rates for the directed edges were chosen in a way that allows to implement local detailed balance\cite{maes2021local}:
\begin{align}
    k_{ij}=k_{ij}^0e^{b\frac{\text{d}(i,N)-\text{d}(j,N)}{2}}.
\end{align}
where $d(i,N)$ and $d(j,N)$ denotes the graph distance from vertex 
$j$ and $i$ to the absorbing target $N$, respectively. The exponential factor allows to tune the bias of the rates: for $b>0$, rates decreasing the distance to $N$ get a higher weight, and for $b<0$, rates increasing the distance to $N$ get a higher weight. The prefactors are assigned at random as $k_{ij}^0\sim U([1-\alpha,1+\alpha])$, to have an average rate of $1$. For all simulations, $\alpha =0.2$ was used. The chosen shape of the rates also means that the detailed balance condition is implemented by choosing $k_{ij}^0=k_{ji}^0$ as a direct consequence of Kolmogorov's criterion. For the general irreversible case, $k_{ij}^0$ and $k_{ji}^0$ are drawn independently for bidirectional transitions.

The eigenvalues of the $N$-th principal minor of the generator matrix were computed by numerically diagonalizing the matrix in python with the numpy linalg.eigvals() function. For the reversible case, the matrix can be symmetrized first using the stationary distribution, allowing to use the more efficient linalg.eigvalsh() function. The largest eigenvalue $\lambda_1^{-1}$ was extracted and the sum of all $\lambda_i^{-1}$ was computed. For the non-reversible case, one has complex eigenvalues that occur in conjugate pairs. For the sum, we used that
\begin{align}
    \sum_{i=1}^N\frac{1}{\lambda_i}=\sum_{i=1}^N\frac{\text{Re}(\lambda_i)}{|\lambda_i|^2},
\end{align}
as shown before, to avoid numerical issues when dealing with complex expressions.



\providecommand{\noopsort}[1]{}\providecommand{\singleletter}[1]{#1}%